\documentclass{article}

\usepackage{arxiv}

\usepackage[utf8]{inputenc} 
\usepackage[T1]{fontenc}    
\usepackage{hyperref}       
\usepackage{url}            
\usepackage{booktabs}       
\usepackage{amsfonts}       
\usepackage{nicefrac}       
\usepackage{microtype}      
\usepackage{lipsum}
\usepackage{graphicx}
\usepackage{epsfig}
\usepackage{subcaption}
\usepackage{xcolor}
\usepackage{pifont}
\usepackage{amsmath,amssymb} 
\usepackage{multirow}
\DeclareMathAlphabet{\pazocal}{OMS}{zplm}{m}{n}

\title{Towards Bidirectional Arbitrary Image Rescaling: Joint Optimization and Cycle Idempotence}

\author{
  Zhihong Pan, Baopu Li \\
  Baidu Research (USA)\\
   \\
   \And
 Dongliang He, Mingde Yao, Wenhao Wu, Tianwei Lin, Xin Li, Errui Ding \\
  Department of Computer Vision Technology (VIS), Baidu Inc.\\
  \\
}

\begin{document}
\newcommand{\etal}{\textit{et al}.}
\newcommand{\cmark}{\ding{51}}%
\newcommand{\xmark}{\ding{55}}%
\newcommand{\red}[1]{\textcolor{red}{#1}}
\newcommand{\bp}[1]{\textcolor{blue}{#1}}
\newcommand{\wt}[1]{\textcolor{white}{#1}}

\maketitle

\begin{abstract}
Deep learning based single image super-resolution models have been widely studied and 
superb results are achieved in upscaling low-resolution images with fixed scale factor and
downscaling degradation kernel.  To improve real world applicability of such models,
there are growing interests to develop models optimized for arbitrary upscaling factors.
Our proposed method is the first to treat arbitrary rescaling, both upscaling and downscaling,
as one unified process.  Using joint optimization of both directions, the proposed model is able to
learn upscaling and downscaling simultaneously and achieve bidirectional arbitrary image
rescaling.  It improves the performance of current arbitrary upscaling models by a large margin 
while at the same time learns to maintain visual perception quality in downscaled images.
The proposed model is further shown to be robust in cycle idempotence test,
free of severe degradations in reconstruction accuracy when the downscaling-to-upscaling
cycle is applied repetitively.  This robustness is beneficial for image rescaling in the wild
when this cycle could be applied to one image for multiple times. 
It also performs well on tests with arbitrary large scales and asymmetric scales,
even when the model is not trained with such tasks.
Extensive experiments are conducted to demonstrate the superior performance of our model.
\end{abstract}

\section{Introduction}
\label{sec:intro}

In real world applications, it is common to rescale an image with arbitrary scale factors,
either scaling up or down, for various purposes like display, storage or transmission.
While recent deep learning based image super-resolution (SR) method have advanced the
performance of image upscaling significantly,
they are mostly optimized for fixed scale factors and known downscaling degradation kernels.
Lately, there are growing interests in SR models that support arbitrary scale factors and great successes
have been achieved, including arbitrary upscaling for scale factors in certain range~\cite{hu_cvpr_2019},
or learning a continuous image representation to resize it at any larger resolution~\cite{chen_cvpr_2021},
or asymmetric arbitrary upscaling where the vertical and horizontal scale factors
could be different~\cite{wang_iccv_2021}.  Like standard SR models, these methods are all
optimized for the unidirectional upscaling process.
In contrast, another line of image rescaling models~\cite{kim_eccv_2018, sun_tip_2020, xiao_eccv_2020}
are developed to optimize the downscaling process together with the inverse upscaling and are
able to improve accuracy on the upscaling task significantly comparing to unidirectional SR models of the same
scale factors.  Currently these bidirectional rescaling models are limited to a specific integer scale
as far as we know.

\begin{figure}[h]
\captionsetup[subfigure]{font=footnotesize, labelformat=empty}
\begin{center}
  \begin{subfigure}[b]{0.095\textwidth}
    \centering
      \includegraphics[width=\textwidth, interpolate=false]{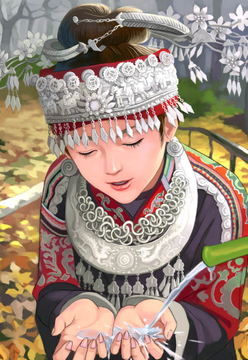}
      \caption{GT}
  \end{subfigure} \hspace*{-0.45em}
  \begin{subfigure}[b]{0.095\textwidth}
    \centering
      \includegraphics[width=\textwidth, interpolate=false]{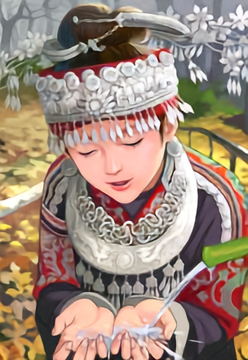}
      \caption{IRN-C1}
  \end{subfigure} \hspace*{-0.45em}
  \begin{subfigure}[b]{0.095\textwidth}
    \centering
      \includegraphics[width=\textwidth, interpolate=false]{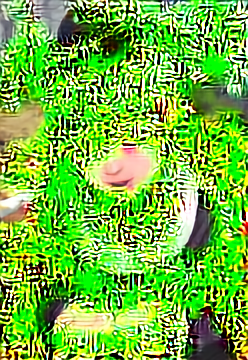}
      \caption{IRN-C2}
  \end{subfigure} \hspace*{-0.45em}
  \begin{subfigure}[b]{0.095\textwidth}
    \centering
      \includegraphics[width=\textwidth, interpolate=false]{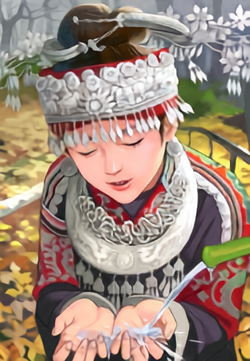}
      \caption{Ours-C1}
  \end{subfigure} \hspace*{-0.45em}
  \begin{subfigure}[b]{0.095\textwidth}
    \centering
      \includegraphics[width=\textwidth, interpolate=false]{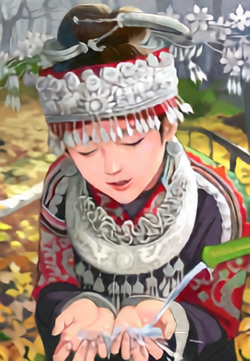}
      \caption{Ours-C5}
  \end{subfigure}
 \end{center}
   \vspace{-3pt}
    \caption{Visual examples of quality degradation from multiple downscaling-to-upscaling cycles: IRN~\cite{xiao_eccv_2020} and ours.}
   \vspace{-3pt}
\label{fig:videm}
\end{figure}

Here we propose a joint optimization process that is able to learn
arbitrary downscaling and arbitrary upscaling simultaneously.
By modeling both downscaling and upscaling as equivalent subpixel splitting
and merging processes and learning through a downscaling-to-upscaling cycle,
the proposed method is able to achieve the best arbitrary upscaling accuracy while maintaining
high perception-quality in downscaled outputs.  An LIIF~\cite{chen_cvpr_2021}-like
subpixel weight function (SVF) and a novel subpixel weight function (SWF) are introduced for
subpixel splitting and merging respectively.  Using ground-truth (GT) image as supervision for
high-resolution (HR) reconstruction,  plus a weak supervision in low-resolution (LR),
the proposed model, jointly optimized for both upscaling and downscaling, is able to greatly advance
performances in arbitrary image rescaling, including very large or asymmetric scales.

In addition, so far as we know, current models are only evaluated for a single application of
the downscaling-to-upscaling cycle and the effects of multiple cycles have never been studied.
Ideally, application of additional cycle should not introduce any further changes beyond
the initial cycle.  This ideal downscaling-to-upscaling process, by definition, is an
idempotent operation.  In other words, for a function $f$ of variable $x$, $f$ is idempotent if
$\forall x, f(f(x)) = f(x)$.  While an ideal idempotent rescaling cycle may not be feasible,
it is desirable to have minimum additional degradation when multiple downscaling-to-upscaling cycles
are applied.  Here a proxy objective is studied for optimization of both reconstruction accuracy and
idempotence, and a cycle idempotence test is introduced to assess the quality of
output image from multiple cycles in comparison to the original GT. 
As shown in Fig.~\ref{fig:videm}, for IRN~\cite{xiao_eccv_2020}, while result from the first cycle
(C1) is of high quality, severe artifacts appear pervasively when the output from C1 is used
as input for C2.  In comparison, results from ours have similar high quality at C1,  and other than minor loss of details,
no visible artifacts even at C5.

In summary, the main contributions of our work include:
\begin{itemize}
\setlength\itemsep{0.05em}
\item[$\bullet$] We are the first to consider bidirectional arbitrary image rescaling, both downscaling and upscaling,
as one unified process and achieve SOTA performance in arbitrary image rescaling via joint optimization.

\item[$\bullet$] A proxy objective that optimize both reconstruction accuracy and idempotence is investigated, and a
newly proposed cycle idempotence test is conducted to demonstrate our method's superior performance
in model robustness after repetitive applications of the downscaling-to-upscaling cycle.

\item[$\bullet$] The proposed method is also applicable to arbitrary asymmetric scales and large out-of-distribution
scales and able to achieve SOTA in both tests.

\end{itemize}

\section{Related Work}
\label{sec:rwork}

\noindent\textbf{{Arbitrary Scale Super-Resolution.}}
Deep learning based image super-resolution have been studied extensively for the last few
years~\cite{dong_eccv_2014, kim_cvpr_2016_2, lim_cvprw_2017, zhang_cvpr_2018, zhang_eccv_2018}, 
and these methods commonly train one model for one fixed scale factor.
Lim \etal~\cite{lim_cvprw_2017} was the first to propose a multi-scale SR model, which shares
one feature learning backbone for different scales but still needs scale specific processing modules
to handle the last step for multiple scales.
Later, Li \etal~\cite{li_eccv_2018} proposed a multi-scale residual network
learn multi-scale spatial features using convolution layers with different kernel sizes.
However, these methods are still limited to a fixed set of integer scale factors.
Inspired by weight prediction techniques in meta-learning~\cite{lemke_air_2015},
Hu \etal~\cite{hu_cvpr_2019} proposed a single Meta-SR model to solve SR at arbitrary scale factors
by predicting weights of convolutional layers for arbitrary scale factors, not limited to
a fixed set of integer ones.
The newest ArbSR~\cite{wang_iccv_2021} proposed a plug-in module to further optimize existing SR models
for arbitrary asymmetric SR where scale factors along horizontal and vertical directions could be different.
These arbitrary SR works are often limited to a fixed maximum scale factor to maintain high performance.
Most recently, Chen \etal~\cite{chen_cvpr_2021} proposed to learn pixel representation
features to replace pixel value features in previous methods.  With
a learned local implicit image function (LIIF), this model is able to predict pixel values at arbitrary
large scales.
Our work extends the idea of LIIF to be applicable for arbitrary downscaling and upscaling at the same time.

\vspace{1pt} \noindent\textbf{{Bidirectional Image Rescaling.}}
As pointed out above, most super-resolution models rely on LR-HR pairs where each LR image is
downscaled from the corresponding HR using frequency-based kernels like Bicubic~\cite{mitchell_siggraph_1988}.
These models are trained for upscaling reconstruction only without taking the image downscaling method
into joint consideration.  To take advantage of the potential
mutually beneficial reinforcement between downscaling and the inverse upscaling,
Kim \etal~\cite{kim_eccv_2018} proposed an auto-encoder framework
to jointly train image downscaling and upscaling together.
Similarly, Sun \etal~\cite{sun_tip_2020} proposed a new content
adaptive-resampler based image downscaling method, which can be jointly trained with
any existing differentiable upscaling (SR) models.
More recently, Xiao \etal~\cite{xiao_eccv_2020} proposed an invertible rescaling net (IRN)
that has set the state-of-the-art (SOTA) for learning based bidirectional image rescaling.
Based on the invertible neural network (INN)~\cite{ardizzone_iclr_2018},
IRN learns to convert HR input to LR
output and an auxiliary latent variable $z$.  By mapping $z$ to a
case-agnostic normal distribution during training, inverse image upscaling is implemented
by randomly sampling $z$ from the normal distribution without need of the case specific $\hat{z}$.
Current methods for bidirectional image rescaling are limited to a fixed integer scale factor
like $\times 4$.  As a contrast, we propose a bidirectional arbitrary rescaling approach in this work.

 \begin{figure*}[t]
 \begin{center}
     \includegraphics[width=\linewidth]{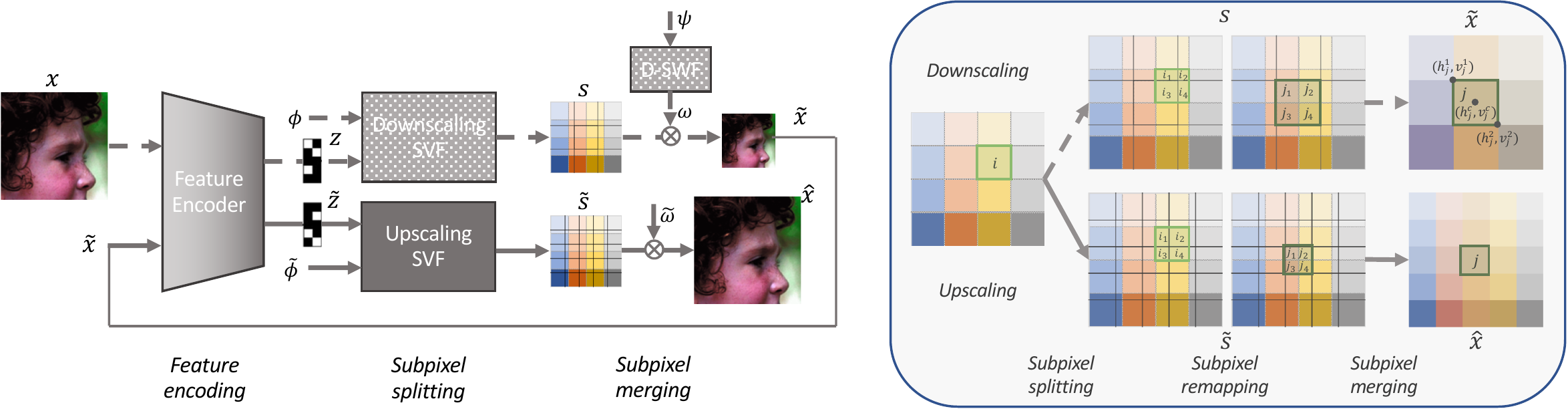}
 \end{center}
 \vspace{-3pt}
 \caption{Proposed framework for a bidirectional arbitrary image rescaling network (BAIRNet) with magnified illustration of the subpixel splitting and merging process.}
 \vspace{-3pt}
 \label{fig:net}
 \end{figure*}

\vspace{1pt} \noindent\textbf{{Idempotent Image Processing.}}
For image processing, there are numerous examples
of idempotent filters such as median filter~\cite{nodes_assp_1982},
cascaded median filters~\cite{haavisto_csc_1991} and basic morphological operations
like opening and closing~\cite{haralick_pami_1987}.
For many image processing application, it is beneficial to have idempotent filters or
processes for various reasons.
In the case of image JPEG compression, an image could be compressed multiple times as
it is not known for sure if an image in the wild is already compressed.
To reflect the importance of repetitive image compression, there
is a specific key feature of multi-generation robustness in the standardization process
of JPEG XS~\cite{descampe_adip_2017}.  Lately, it has been discovered that
there is model instability issue in successive deep image compression that results
in severe visual artifacts~\cite{kim_acmmm_2020}.
Here we will specifically study cycle idempotence of image SR and
rescaling models after repetitive applications of the downscaling-to-upscaling process.

\vspace{1pt} \noindent\textbf{{Weak Supervision in Low Level Vision.}}
Weak supervision in a branch of supervised learning where supervision signals, like labels
for image classification, are from either imprecise or noisy sources.
While it has been widely studied in high level tasks like object detection~\cite{bilen_cvpr_2015}
and semantic segmentation~\cite{khoreva_cvpr_2017}, its applications in low level vision tasks
like image restoration and reconstruction, however, are not fully addressed.
For image SR where LR-HR pairs are needed, LR images are often synthesized from HR
using bicubic interpolation and they could be imprecise comparing to real world LR images.
There have been efforts to co-collect true LR and HR images to build
real-world SR dataset like RealSR~\cite{cai_iccv_2019} and DRealSR~\cite{wei_eccv_2020}.  However,
these pairs are subject to registration imprecision and/or local motion blur as LR and HR are taken
sequentially using different lenses.
For newest bidirectional rescaling models~\cite{kim_eccv_2018, sun_tip_2020, xiao_eccv_2020} that learn
downscaling and upscaling jointly, although accuracy in inverse upscaling is the primary goal,
a weakly supervised learning for downscaling is still needed.  While the LR image from bicubic interpolation
has been previously used as the downscaling reference, new forms of weak supervision are investigated in our study.

\section{Proposed Method}
\label{sec:method}

\subsection{Bidirectional Arbitrary Image Rescaling}

The proposed framework for a bidirectional arbitrary image rescaling network (BAIRNet) is illustrated in Fig.~\ref{fig:net}.
It is a bidirectional process, including downscaling to convert the GT image $x$ to an LR image $\tilde{x}$,
and upscaling to restore a HR image $\hat{x}$ from $\tilde{x}$.
As illustrated on the left, each of the two directions consists of the same three steps:
feature encoding, subpixel splitting and subpixel merging.  These steps are denoted as
 \begin{equation}
 \begin{split}
\!\tilde{x}&\!=\!f_m(s,\!\omega)\!=\!f_m(f^D_s(z,\!\phi),\!\omega)\!=\!f_m(f^D_s(f_e(x),\!\phi),\!\omega)\\
\!\hat{x}&\!=\!f_m(\tilde{s},\!\tilde{\omega})\!=\!f_m(f^U_s(\tilde{z},\!\tilde{\phi}),\!\tilde{\omega})\!=\!f_m(f^U_s(f_e(\tilde{x}),\!\tilde{\phi}),\!\tilde{\omega})
 \end{split}
 \label{eq:func}
 \end{equation}
for downscaling and upscaling respectively.  For the three steps,
feature encoding $f_e$ is shared for both directions and subpixel merging $f_m$ is the same definition too.
For subpixel splitting though, different subpixel value function (SVF) are trained, denoted as $f^D_s$ and $f^U_s$ respectively.
For variables, $z/\tilde{z}$ are feature vectors, $s/\tilde{s}$ are subpixel values,
$\phi/\tilde{\phi}$ are subpixel coordinates, and $\omega/\tilde{\omega}$
are subpixel weights used in  merging.  Each pair has the same definition
and \space$\tilde{}$\space\space is used to differentiate upscaling from downscaling.

To illustrate the subpixel splitting and merging process,
a magnified illustration is included in Fig.~\ref{fig:net} on the right.
One pixel $i$ of the input is split to one or more subpixels first, and after remapping,
a new set of subpixels is merged to one pixel $j$ in the rescaled image.
A subpixel $k$ in the intermediate step is defined as one rectangle in the image space that reside wholly
inside one original pixel as well as inside one rescaled pixel,
and its boundaries are aligned with the boundaries of input and/or rescaled pixels.
Here we use $p_i$, $s_k$ and $r_j$ to represent values of pixels $i$, $j$ and $k$ respectively.
In the illustrated downscaling example, pixel $i$ is split to four subpixels and this mapping relationship
is denoted as $\mathbb{P}_i = \{i_1, i_2, i_3, i_4\}$.  To merge subpixels to rescaled pixels, a remapping
process is needed to associate groups of subpixels corresponding to rescaled pixels.  In the example in
Fig.~\ref{fig:net}, output pixel $j$ is merged from 4 pixels and it is denoted as
$\mathbb{R}_j = \{j_1, j_2, j_3, j_4\}$.  Lastly, as illustrated in the downscaled pixel $j$,
its central, top-left and bottom-right coordinates are denoted as $(h^c_j, v^c_j)$, 
$(h^1_j, v^1_j)$ and $(h^2_j, v^2_j)$ respectively.

For input pixel $i$ and subpixel $k, k \in \mathbb{P}_i$, defining $\phi^i_k$ as relative coordinates
of $k$ in reference to $i$: $\phi^i_k = (h^1_k-h^c_i, v^1_k-v^c_i, h^2_k-h^c_i, v^2_k-v^c_i)$,
the process to predict subpixel values $s_k$ during subpixel-splitting is denoted as
 \begin{equation}
 s_k = f_s(z_i, \phi^i_k)
 \end{equation}
where $f_s$ is the SVF function and $z_i$ is the feature vector of pixel $i$.
This process is the same for both downscaling and upscaling but separate SVFs are trained and denoted
differently in Eq.~\ref{eq:func} for distinction.

The value of pixel $j$ at subpixel-merging is computed as
 \begin{equation}
 r_j = \textstyle\sum_{k \in \mathbb{R}_j} \omega^j_k s_k / \textstyle\sum_{k \in \mathbb{R}_j} \omega^j_k
 \end{equation}
where $\omega^j_k$ is the weight of subpixel $k$ during merging of pixel $j$.
For the subpixel merging weights in upscaling, $\tilde{\omega}^j_k$ is simply defined as
the area of subpixel $k$.
As the majority of upscaled pixel $j$ consists of just one subpixel $k$, and the others have either 2 or 4,
the area based weights are sufficient to represent the significance of each subpixel.
While in the case of downscaling, each pixel $j$ may include a large number of subpixels and the impact
of each subpixel should dependent on both its size and location.
Here we propose a subpixel weight function (SWF) module to learn
the subpixel weights for merging during the end-to-end training, denoted as $\omega^j_k = f_w(\psi^j_k)$.
Similar to $\phi^i_k$, $\psi^j_k$ is defined as $(h^1_k-h^c_j, v^1_k-v^c_j, h^2_k-h^c_j, v^2_k-v^c_j)$.

While this framework has some resemblance with two prior works, that is, IRN~\cite{xiao_eccv_2020} and
LIIF~\cite{chen_cvpr_2021}, there are some substantial differences between our proposed and the previous
two.  First, IRN is limited to one fixed integer scale per trained model.  Although
it is also trained to optimize both downscaling and upscaling together, it is based on
an invertible network which uses forward and backward inferences for downscaling and upscaling respectively.
In contrast, ours is utilizing the same three-step process for both directions, and only one model
is needed to handle arbitrary scales.
IRN also samples auxiliary latent variables randomly during the backward upscaling process which brings
uncertainty and causes severe artifacts in cycle idempotence tests.  Comparing to LIIF, our model consolidates
the downscaling and upscaling process to be utilize similar implicit functions for both arbitrary downscaling
and upscaling.  As a result, it can be trained for bidirectional arbitrary rescaling and leads to great
improvements in performance.  Lastly, asymmetric scales are not studied in LIIF.

\subsection{Idempotent Image Rescaling}
\label{sec:iir}
The rescaling cycle defined in Eq.~\ref{eq:func} can be simplified as $\hat{x} = f(x)$.
Without considering constraints in LR, the primary goal to optimize this cycle is 
to minimize its reconstruction loss, but it is also desirable to learn an idempotent one.
These two objectives could be defined separately as
 \begin{equation}
 \begin{split}
 f & = \textstyle{\min_{\zeta}} \mathcal{L} (x, f_\zeta(x)) \\
 f & = \textstyle{\min_{\eta}} \mathcal{L} (f_\eta(x), f_\eta(f_\eta(x)))
 \end{split}
 \end{equation}
As these two objectives may conflict, an empirical proxy objective is proposed to learn a compromise
between the two.  In practice, the model is trained to minimize reconstruction error after $n$ cycles,
described as
 \begin{equation}
 f = \textstyle{\min_{\theta}} \mathcal{L} (x, f^n_\theta(x)), n \in [1, N]
 \label{eq:optm}
 \end{equation}
where $f^n_\theta$ means $f_\theta$ is applied $n$ times.
When $N$ is set as 1, this proxy objective is equivalent to the primary task of 1-cycle reconstruction.
In our experiments, different $N$ are investigated to compare the trade-off between two objectives:
reconstruction accuracy and cycle idempotence.

\begin{table*}[ht!]
	\centering
	\footnotesize
	\setlength{\tabcolsep}{5pt}
\vspace{-3pt}
	\caption{Quantitative comparison of SOTA SR and rescaling methods with the best two results highlighted in \red{red} and \bp{blue} respectively (methods in \textbf{bold} require multiple models and additional interpolations to achieve arbitrary scales).} \label{tab:main}
\vspace{-3pt}
	\begin{tabular}{cccccccc} 
		\hline
		{Method} & {Scale} & {Set5} & {Set14} & {BSD100} & {Urban100} & {Manga109} & {DIV2K}\\
		\hline \hline
		Bicubic       & {$\times$1.5} & 36.75/0.9611 & 32.86/0.9268 & 32.16/0.9133 & 29.49/0.9095 & 34.79/0.9707 & 33.95/0.9416\\
		\textbf{RCAN}~\cite{zhang_eccv_2018}& {$\times$1.5} & 40.97/0.9767 & 37.05/0.9578 & 35.59/0.9516 & 35.93/0.9660 & 42.33/0.9889 & 38.47/0.9701\\
		Meta-SR~\cite{hu_cvpr_2019}& {$\times$1.5} & 41.47/0.9785 & 37.52/0.9601 & 35.86/0.9543 & 36.91/0.9696 & \bp{43.17/0.9904} & 38.88/0.9718\\
		LIIF~\cite{chen_cvpr_2021}& {$\times$1.5} & 41.23/0.9774 & 37.37/0.9591 & 35.76/0.9536 & 36.70/0.9684 & 42.84/0.9894 & 38.82/0.9717\\
		ArbSR~\cite{wang_iccv_2021}& {$\times$1.5} & 41.47/0.9786 & 37.51/0.9603 & 35.86/0.9547 & \bp{36.92/0.9697} & 43.12/0.9904 & 38.84/0.9719\\
		\hline
		\textbf{CAR}~\cite{sun_tip_2020} & {$\times$1.5} & 40.50/0.9763 & 37.08/0.9596 & 35.72/0.9535 & 34.70/0.9635 & 40.90/0.9881 & 37.93/0.9683\\
		\textbf{IRN}~\cite{xiao_eccv_2020}& {$\times$1.5} & \bp{43.55/0.9891} & \bp{39.52/0.9795} & \bp{39.28/0.9833} & 36.52/0.9811 & 42.64/0.9936 & \bp{40.18/0.9838}\\
	BAIRNet$^\dagger$ & {$\times$1.5} & \red{47.13/0.9849} & \red{43.12/0.9760} & \red{46.63/0.9959} & \red{44.01/0.9946} & \red{45.49/0.9948} & \red{44.99/0.9920}\\
		\hline \hline
		Bicubic       & {$\times$2.5} & 31.76/0.8983 & 28.52/0.8196 & 28.13/0.7853 & 25.43/0.7837 & 28.56/0.8954 & 29.40/0.8505\\
		\textbf{RCAN}~\cite{zhang_eccv_2018}& {$\times$2.5} & 36.05/0.9436 & 31.69/0.8815 & 30.47/0.8508 & 30.42/0.8990 & 36.59/0.9634 & 32.72/0.9079\\
		Meta-SR~\cite{hu_cvpr_2019}& {$\times$2.5} & 36.18/0.9441 & 31.90/0.8814 & 30.47/0.8508 & 30.57/0.9003 & 36.55/0.9639 & 32.77/0.9086\\
		LIIF~\cite{chen_cvpr_2021}& {$\times$2.5} & 35.98/0.9434 & 31.64/0.8813 & 30.45/0.8510 & 30.42/0.8992 & 36.39/0.9630 & 32.78/0.9091\\
		ArbSR~\cite{wang_iccv_2021}& {$\times$2.5} & 36.21/0.9448 & 31.99/0.8830 & 30.51/0.8536 & 30.68/0.9027 & 36.67/0.9646 & 32.77/0.9093\\
		\hline
		\textbf{CAR}~\cite{sun_tip_2020}& {$\times$2.5} & 37.33/0.9548 & 33.78/0.9169 & 32.53/0.9020 & 32.19/0.9301 & 37.63/0.9717 & 34.32/0.9310\\
		\textbf{IRN}~\cite{xiao_eccv_2020}& {$\times$2.5} & \bp{39.78/0.9742} & \bp{36.39/0.9553} & \bp{35.56/0.9542} & \bp{33.99/0.9589} & \bp{39.33/0.9836} & \bp{36.60/0.9607}\\
	BAIRNet$^\dagger$ & {$\times$2.5} & \red{40.11/0.9664} & \red{36.62/0.9469} & \red{36.29/0.9563} & \red{36.62/0.9679} & \red{40.26/0.9830} & \red{37.46/0.9627}\\
		\hline \hline
		Bicubic       & {$\times$3.5} & 29.30/0.8374 & 26.52/0.7362 & 26.50/0.7003 & 23.70/0.6935 & 25.83/0.8203 & 27.38/0.7802\\
		\textbf{RCAN}~\cite{zhang_eccv_2018}& {$\times$3.5} & 33.47/0.9138 & 29.24/0.8141 & 28.42/0.7731 & 27.61/0.8348 & 32.74/0.9328 & 30.13/0.8511\\
		Meta-SR~\cite{hu_cvpr_2019}& {$\times$3.5} & 33.59/0.9146 & 29.60/0.8140 & 28.42/0.7728 & 27.71/0.8356 & 32.75/0.9337 & 30.18/0.8524\\
		LIIF~\cite{chen_cvpr_2021}& {$\times$3.5} & 33.41/0.9133 & 29.20/0.8131 & 28.39/0.7714 & 27.60/0.8334 & 32.60/0.9324 & 30.16/0.8517\\
		ArbSR~\cite{wang_iccv_2021}& {$\times$3.5} & 33.63/0.9149 & 29.58/0.8147 & 28.41/0.7744 & 27.69/0.8360 & 32.84/0.9339 & 30.14/0.8518\\
		\hline
		\textbf{CAR}~\cite{sun_tip_2020}& {$\times$3.5} & 34.98/0.9303 & 31.38/0.8643 & 30.14/0.8326 & 29.97/0.8871 & 35.00/0.9507 & 31.88/0.8865\\
		\textbf{IRN}~\cite{xiao_eccv_2020}& {$\times$3.5} & \red{37.12/0.9546} & \red{33.65/0.9196} & \red{32.54/0.9047} & \bp{31.84/0.9277} & \bp{36.86/0.9690} & \bp{33.84/0.9281}\\
	BAIRNet$^\dagger$ & {$\times$3.5} & \bp{36.85/0.9472} & \bp{32.97/0.9074} & \bp{32.36/0.8986} & \red{32.71/0.9338} & \red{36.98/0.9671} & \red{33.87/0.9266}\\
		\hline	\end{tabular}
\label{tab:scale}
\vspace{-1pt}
\end{table*}

\subsection{Weak Supervision in LR}
Considering the multi-cycle optimization as in Eq.~\ref{eq:optm} and the need
to generate visually coherent LR images, the overall loss for training our model is defined as
 \begin{equation}
 \mathcal{L} = \lambda_1 \mathcal{L}_{rec}(x, \hat{x}^n) + \lambda_2 \mathcal{L}_{ref}(x, \tilde{x}^n)
 \end{equation}
where $\mathcal{L}_{rec}$ is the reconstruction loss for HR, $\mathcal{L}_{ref}$ is the reference
loss in LR, and $\tilde{x}^n$ and $\hat{x}^n$ are the LR and HR outputs after n-cycles respectively.
Although it is possible to train it fully self-supervised by setting $\lambda_2 = 0$,
it will lead to visually non-meaningful $\tilde{x}$ due to random initialization.
In previous methods~\cite{kim_eccv_2018, sun_tip_2020, xiao_eccv_2020}, a $L_{2}$ reference loss
$L_2(\tilde{x}, \bar{x})$, where $\bar{x}$ is the LR reference
downsampled from $x$ using Bicubic~\cite{mitchell_siggraph_1988} method, is used as an imprecise
supervision.  In contrast to previous methods,
various strategies, like reducing $\lambda_2$ to 0 at later stage of training,
or calculating $\mathcal{L}_{ref}$ from the mean values of each color channel instead of per-pixel,
are explored in our study to demonstrate the advantage of weak supervision in LR.

\section{Experiments}
\label{sec:exp}

\subsection{Data and Settings}
\label{sec:set}
For fair comparison with previous works like LIIF and IRN, the same $800$ HR images from 
DIV2K~\cite{agustsson_ntire_2017} are used for training.
For quantitative evaluation, we use HR images of five commonly used benchmark datasets,
including Set5 \cite{bevilacqua_bmvc_2012}, Set14~\cite{zeyde_iccs_2010},
BSD100~\cite{martin_iccv_2001}, Urban100~\cite{huang_cvpr_2015} and Manga109~\cite{huang_cvpr_2015},
plus $100$ HR images from the DIV2K validation set.  
Following previous practices like LIIF, we take the peak noise-signal ratio (PSNR) and
SSIM~\cite{wang_tip_2004} on the luminance channel for the 5 benchmark sets,
but use the same metrics in RGB color space for DIV2K validation set.

The are $B$ $200 \times 200$ input HR patches in one batch, each with a
random scale sampled from a uniform distribution of $\pazocal{U}(1, 4)$.
For individual modules, we use RDN~\cite{zhang_cvpr_2018} minus its upsampling module as the feature encoder,
which generates a feature map with the same size as the input image.
For both downscaling and upscaling SVF, a 5-layer MLP with ReLU activation and hidden dimensions of 256
is used.  For the downscaling SWF, a 5-layer MLP with hidden dimensions of 16 is used.
With a batch size of 8, all models are trained using Adam~\cite{kingma_arxiv_2014} optimizer.
In order to conduct ablation studies efficiently, a pretrained model is generated after
500 epochs, 300 iteration each, from an initial learning rate of $10^{-4}$.
The learning rate decays by half after every 100 epochs.  For this stage, 
$\mathcal{L}_{rec}$ is set as a pixel-level L1 loss and $\mathcal{L}_{ref}$ is set as L2, and no SWF
module is included.  The pretrained one is further trained for 500 epochs get the base model BAIRNet,
with downscaling SWF included, and $\mathcal{L}_{ref}$ is set as L2 for the
mean pixel value per color channel.  $\lambda_1$ and $\lambda_2$ are set as 1 unless specified otherwise.
BAIRNet is further fine-tuned for 200 epochs using the proxy objective as defined in Eq.~\ref{eq:optm},
where $N$ is set as 3 and the final model is denoted as BAIRNet$^\dagger$ with $\dagger$ used for distinction.

\begin{figure*}[h!]
\captionsetup[subfigure]{font=footnotesize, labelformat=empty}
\begin{center}
  \begin{subfigure}[b]{0.135\textwidth}
    \centering
      \includegraphics[width=\textwidth, interpolate=false]{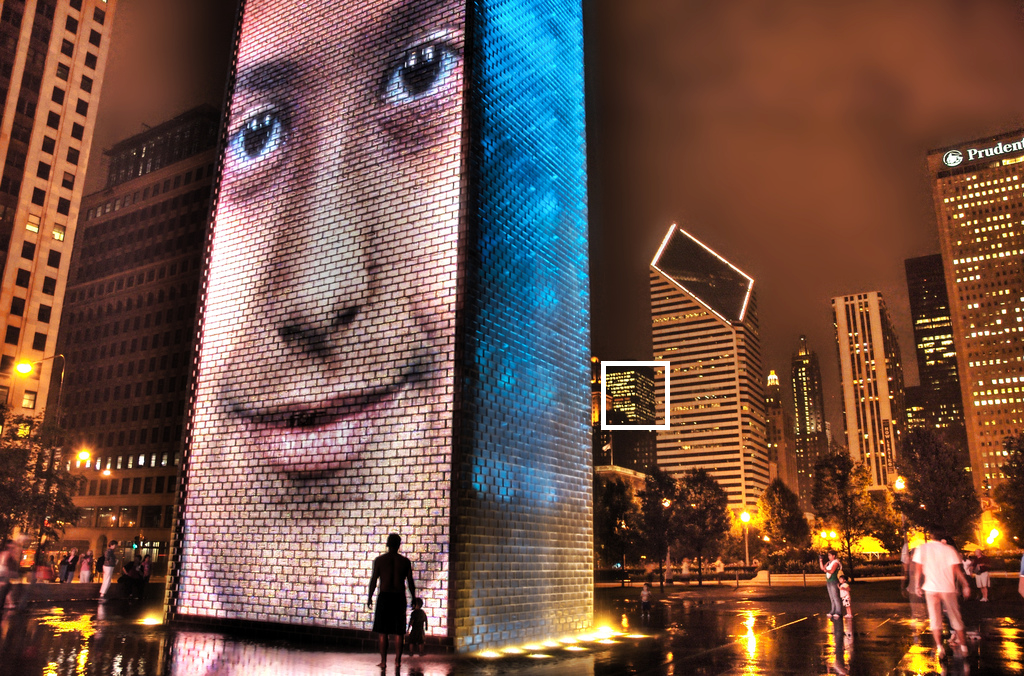}
  \end{subfigure} \hspace*{-0.4em}
  \begin{subfigure}[b]{0.09\textwidth}
    \centering
      \includegraphics[width=\textwidth, interpolate=false]{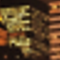}
  \end{subfigure} \hspace*{-0.4em}
  \begin{subfigure}[b]{0.09\textwidth}
    \centering
      \includegraphics[width=\textwidth, interpolate=false]{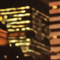}
  \end{subfigure} \hspace*{-0.4em}
  \begin{subfigure}[b]{0.09\textwidth}
    \centering
      \includegraphics[width=\textwidth, interpolate=false]{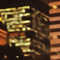}
  \end{subfigure} \hspace*{-0.4em}
  \begin{subfigure}[b]{0.09\textwidth}
    \centering
      \includegraphics[width=\textwidth, interpolate=false]{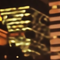}
  \end{subfigure} \hspace*{-0.4em}
  \begin{subfigure}[b]{0.09\textwidth}
    \centering
      \includegraphics[width=\textwidth, interpolate=false]{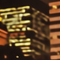}
  \end{subfigure} \hspace*{-0.4em}
  \begin{subfigure}[b]{0.09\textwidth}
    \centering
      \includegraphics[width=\textwidth, interpolate=false]{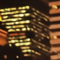}
  \end{subfigure} \hspace*{-0.4em}
  \begin{subfigure}[b]{0.09\textwidth}
    \centering
      \includegraphics[width=\textwidth, interpolate=false]{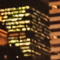}
  \end{subfigure} \hspace*{-0.4em}
  \begin{subfigure}[b]{0.09\textwidth}
    \centering
      \includegraphics[width=\textwidth, interpolate=false]{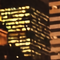}
  \end{subfigure}
  \hspace*{0.8em}
  \rotatebox[origin=l]{90}{\makebox[0.08\textwidth]{$\scriptstyle \times 2.5$}}

  \hspace*{2em}
  \begin{subfigure}[b]{0.09\textwidth}
    \centering
      \includegraphics[width=\textwidth, interpolate=false]{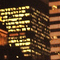}
  \end{subfigure} \hspace*{-0.4em}
  \begin{subfigure}[b]{0.09\textwidth}
    \centering
      \includegraphics[width=\textwidth, interpolate=false]{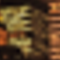}
  \end{subfigure} \hspace*{-0.4em}
  \begin{subfigure}[b]{0.09\textwidth}
    \centering
      \includegraphics[width=\textwidth, interpolate=false]{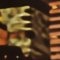}
  \end{subfigure} \hspace*{-0.4em}
  \begin{subfigure}[b]{0.09\textwidth}
    \centering
      \includegraphics[width=\textwidth, interpolate=false]{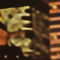}
  \end{subfigure} \hspace*{-0.4em}
  \begin{subfigure}[b]{0.09\textwidth}
    \centering
      \includegraphics[width=\textwidth, interpolate=false]{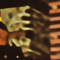}
  \end{subfigure} \hspace*{-0.4em}
  \begin{subfigure}[b]{0.09\textwidth}
    \centering
      \includegraphics[width=\textwidth, interpolate=false]{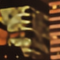}
  \end{subfigure} \hspace*{-0.4em}
  \begin{subfigure}[b]{0.09\textwidth}
    \centering
      \includegraphics[width=\textwidth, interpolate=false]{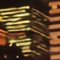}
  \end{subfigure} \hspace*{-0.4em}
  \begin{subfigure}[b]{0.09\textwidth}
    \centering
      \includegraphics[width=\textwidth, interpolate=false]{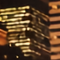}
  \end{subfigure} \hspace*{-0.4em}
  \begin{subfigure}[b]{0.09\textwidth}
    \centering
      \includegraphics[width=\textwidth, interpolate=false]{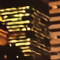}
  \end{subfigure}
  \hspace*{0.8em}
  \rotatebox[origin=l]{90}{\makebox[0.08\textwidth]{$\scriptstyle \times 3.5$}}

  \begin{subfigure}[b]{0.135\textwidth}
    \centering
      \includegraphics[width=\textwidth, interpolate=false]{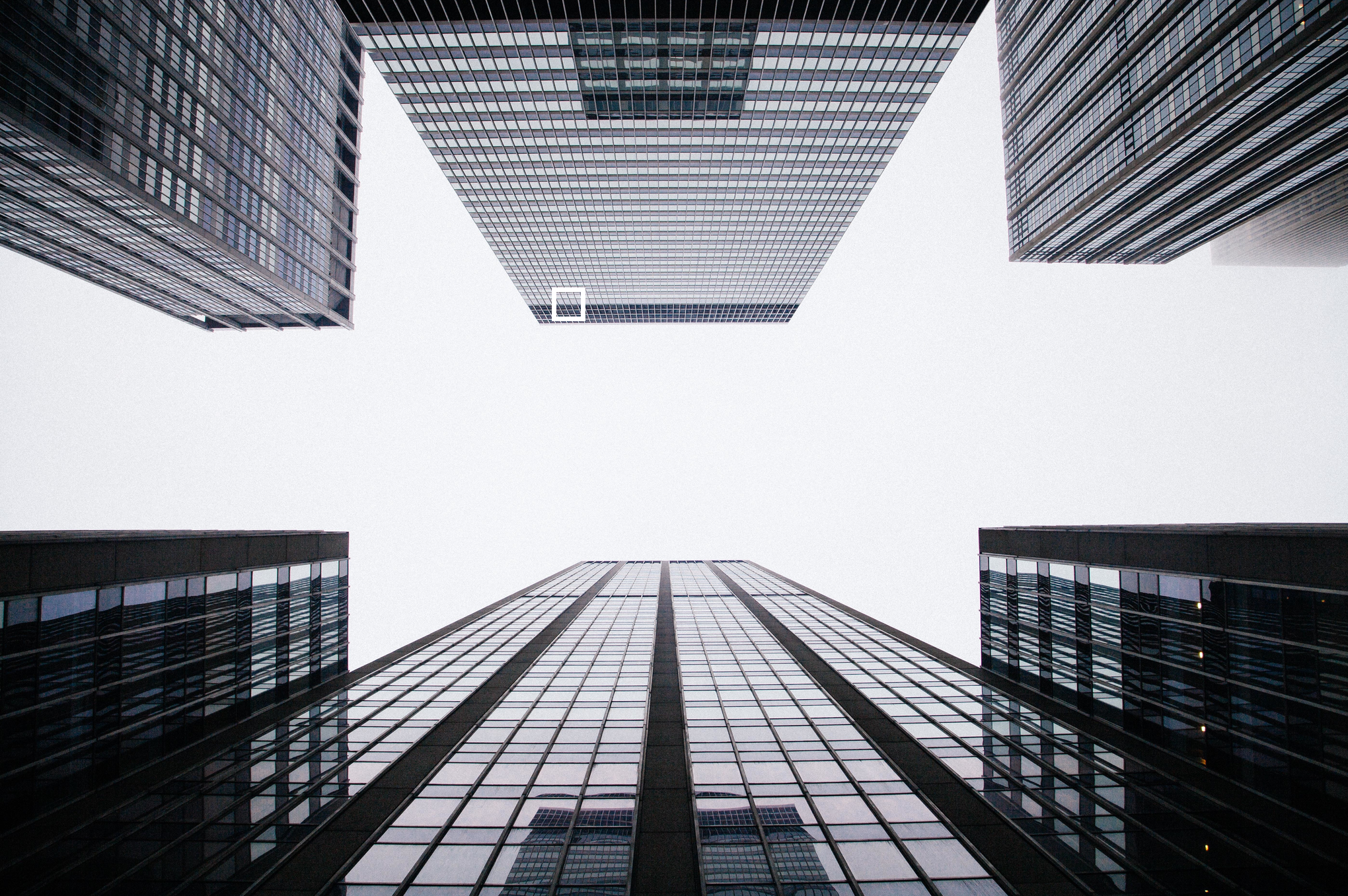}
  \end{subfigure} \hspace*{-0.4em}
  \begin{subfigure}[b]{0.09\textwidth}
    \centering
      \includegraphics[width=\textwidth, interpolate=false]{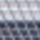}
  \end{subfigure} \hspace*{-0.4em}
  \begin{subfigure}[b]{0.09\textwidth}
    \centering
      \includegraphics[width=\textwidth, interpolate=false]{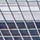}
  \end{subfigure} \hspace*{-0.4em}
  \begin{subfigure}[b]{0.09\textwidth}
    \centering
      \includegraphics[width=\textwidth, interpolate=false]{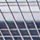}
  \end{subfigure} \hspace*{-0.4em}
  \begin{subfigure}[b]{0.09\textwidth}
    \centering
      \includegraphics[width=\textwidth, interpolate=false]{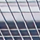}
  \end{subfigure} \hspace*{-0.4em}
  \begin{subfigure}[b]{0.09\textwidth}
    \centering
      \includegraphics[width=\textwidth, interpolate=false]{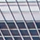}
  \end{subfigure} \hspace*{-0.4em}
  \begin{subfigure}[b]{0.09\textwidth}
    \centering
      \includegraphics[width=\textwidth, interpolate=false]{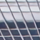}
  \end{subfigure} \hspace*{-0.4em}
  \begin{subfigure}[b]{0.09\textwidth}
    \centering
      \includegraphics[width=\textwidth, interpolate=false]{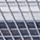}
  \end{subfigure} \hspace*{-0.4em}
  \begin{subfigure}[b]{0.09\textwidth}
    \centering
      \includegraphics[width=\textwidth, interpolate=false]{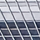}
  \end{subfigure}
  \hspace*{0.8em}
  \rotatebox[origin=l]{90}{\makebox[0.08\textwidth]{$\scriptstyle \times 2.5$}}
  
  \hspace*{2em}
  \begin{subfigure}[b]{0.09\textwidth}
    \centering
      \includegraphics[width=\textwidth, interpolate=false]{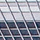}
      \caption{GT\wt{$^\dagger$}}
  \end{subfigure} \hspace*{-0.4em}
  \begin{subfigure}[b]{0.09\textwidth}
    \centering
      \includegraphics[width=\textwidth, interpolate=false]{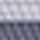}
      \caption{Bicubic\wt{$^\dagger$}}
  \end{subfigure} \hspace*{-0.4em}
  \begin{subfigure}[b]{0.09\textwidth}
    \centering
      \includegraphics[width=\textwidth, interpolate=false]{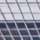}
      \caption{\textbf{RCAN}~\cite{zhang_eccv_2018}\wt{$^\dagger$}}
  \end{subfigure} \hspace*{-0.4em}
  \begin{subfigure}[b]{0.09\textwidth}
    \centering
      \includegraphics[width=\textwidth, interpolate=false]{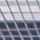}
      \caption{MetaSR~\cite{hu_cvpr_2019}\wt{$^\dagger$}}
  \end{subfigure} \hspace*{-0.4em}
  \begin{subfigure}[b]{0.09\textwidth}
    \centering
      \includegraphics[width=\textwidth, interpolate=false]{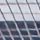}
      \caption{LIIF~\cite{chen_cvpr_2021}\wt{$^\dagger$}}
  \end{subfigure} \hspace*{-0.4em}
  \begin{subfigure}[b]{0.09\textwidth}
    \centering
      \includegraphics[width=\textwidth, interpolate=false]{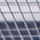}
      \caption{ArbSR~\cite{wang_iccv_2021}\wt{$^\dagger$}}
  \end{subfigure} \hspace*{-0.4em}
  \begin{subfigure}[b]{0.09\textwidth}
    \centering
      \includegraphics[width=\textwidth, interpolate=false]{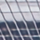}
      \caption{\textbf{CAR}~\cite{sun_tip_2020}\wt{$^\dagger$}}
  \end{subfigure} \hspace*{-0.4em}
  \begin{subfigure}[b]{0.09\textwidth}
    \centering
      \includegraphics[width=\textwidth, interpolate=false]{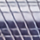}
      \caption{\textbf{IRN}~\cite{xiao_eccv_2020}\wt{$^\dagger$}}
  \end{subfigure} \hspace*{-0.4em}
  \begin{subfigure}[b]{0.09\textwidth}
    \centering
      \includegraphics[width=\textwidth, interpolate=false]{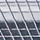}
      \caption{BAIRNet$^\dagger$}
  \end{subfigure}
  \hspace*{0.8em}
  \rotatebox[origin=l]{90}{\makebox[0.1\textwidth]{$\scriptstyle \times 3.5$}}
 \end{center}
 \vspace{-3pt}
    \caption{Visual examples of arbitrary rescaling from Urban100 and DIV2K at two scales: $\times 2.5$ and $\times 3.5$ (Best viewed for online version).}
 \vspace{-3pt}
\label{fig:img}
\end{figure*}

\subsection{Arbitrary Rescaling Performance}
\begin{figure}[t]
 \begin{center}
     \includegraphics[width=0.5\linewidth]{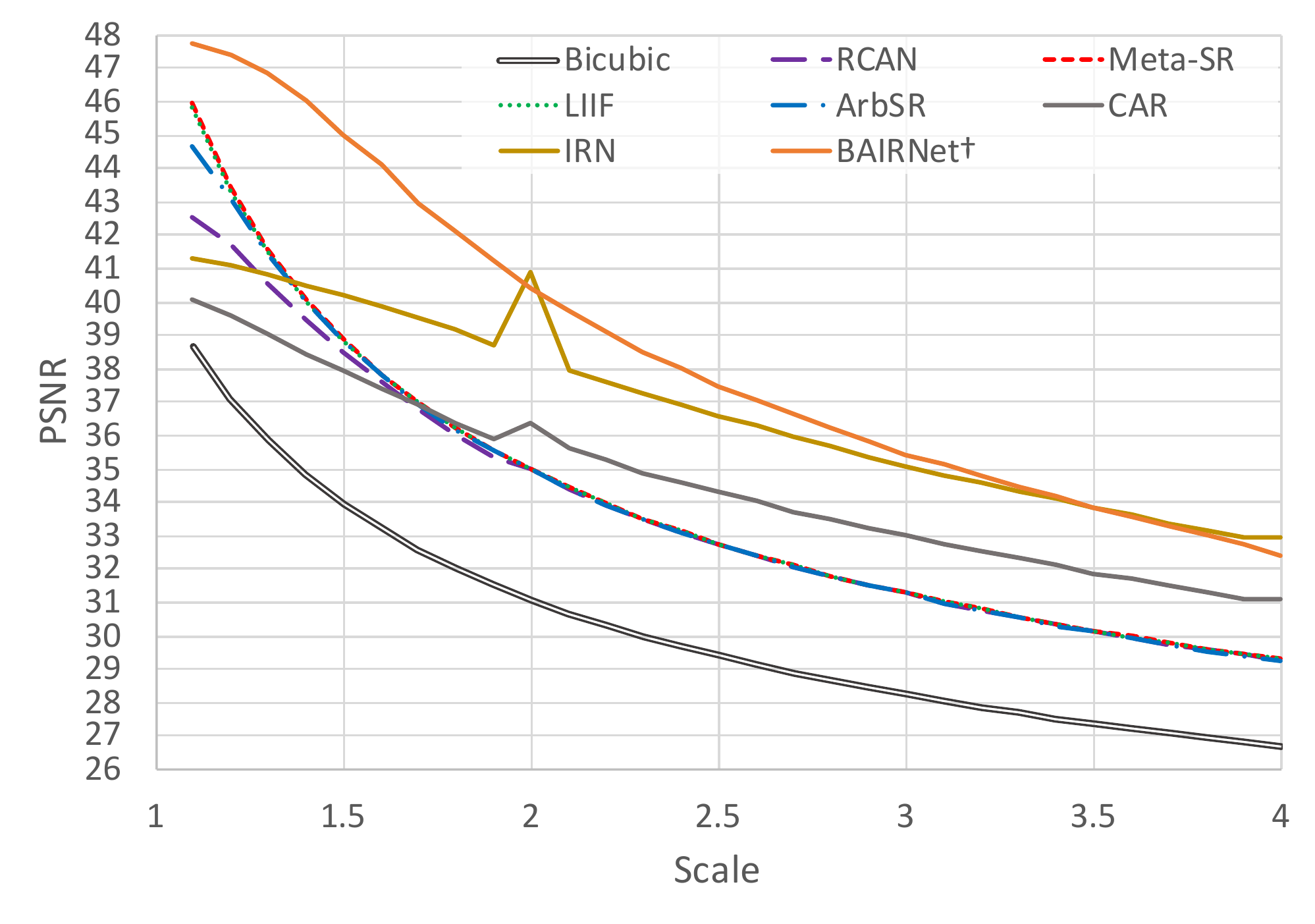}
 \end{center}
 \vspace{-3pt}
 \caption{Cross-scale performance comparison for arbitrary rescaling testing ($\times 1.1-\times 4$) on DIV2K validation set.}
 \vspace{-3pt}
 \label{fig:scale}
 \end{figure}

\vspace{-2pt}
To assess the performance of our proposed method for arbitrary rescaling, we compare rescaled HR images
using a set of arbitrary scales.
For each fixed scale, the resolution of LR images are kept the same for all methods for fair comparison.
For models trained for integer scales only, like RCAN and IRN, evaluation on arbitrary scales is
implemented as upscaling LR using the closest oversampled integer scale (use $\times 3$ for any
scales between 2 and 3) before resampling using bicubic interpolation to target size.
For bidirectional CAR~\cite{sun_tip_2020} and IRN, HR inputs are also pre-upsampled accordingly.
As listed in Table~\ref{tab:scale}, PSNR and SSIM results from three scales ($\times 1.5/2.5/3.5$)
are compared.  It shows that our BAIRNet$^\dagger$ outperforms others by a comfortable margin for 
$\times 1.5$ and $\times 2.5$, and it is the best in $\times 3.5$ tests for the 3 large test sets out of 6 while
trailing slightly behind IRN for the other 3.
Visually as shown in Fig.~\ref{fig:img},
bidirectional methods like IRN and ours are the best overall.
Between the two, IRN is more blurry in $\times 2.5$ and its color is off in the second
example of $\times 3.5$.
Results of continuous scales between $\times 1.1$ and $\times 4$ (sampled every 0.1)
are also illustrated in Fig.~\ref{fig:scale} to compare our model with others.
For arbitrary upscaling models like Meta-SR, LIIF and ArbSR, they are essentially equivalent to RCAN and with each
other for scales above $\times 2$.  Bidirectional models CAR and IRN improves performances in larger scales greatly
but their performances suffer at small arbitrary scales.
BAIRNet$^\dagger$ is clearly the best overall, at the top for all scales except trailing slightly behind
IRN for scales above $\times 3.5$, plus the spike at $\times 2$, where no extra interpolation needed
is needed for IRN. 

\subsection{Cycle Idempotence}
A cycle idempotence test is defined as the assessment of $\mathcal{L}(x, f^n(x))$
for different number of cycles, where $f^n$ means the rescaling cycle
$f$ is applied $n$ times.  Here we use the PSNR value in place of
$\mathcal{L}$ for test assessment.  For the first set of test, defined as closed test,
the downscaling function is fixed as the one best matches its upscaling one.
So for RCAN, Meta-SR, ArbSR and LIIF, matlab\_imresize~\cite{matlab_imresize} is used
for its equivalence with the Matlab one.
For other bidirectionally trained models, their own corresponding downscaling
process are applied respectively.
For the open test, which means the downscaling is set freely, cv2.resize with INTER\_AREA
interpolation is picked for its wide application, and it is used for all methods for fair comparison.

 \begin{figure}[h]
 \begin{center}
     \includegraphics[width=0.7\linewidth]{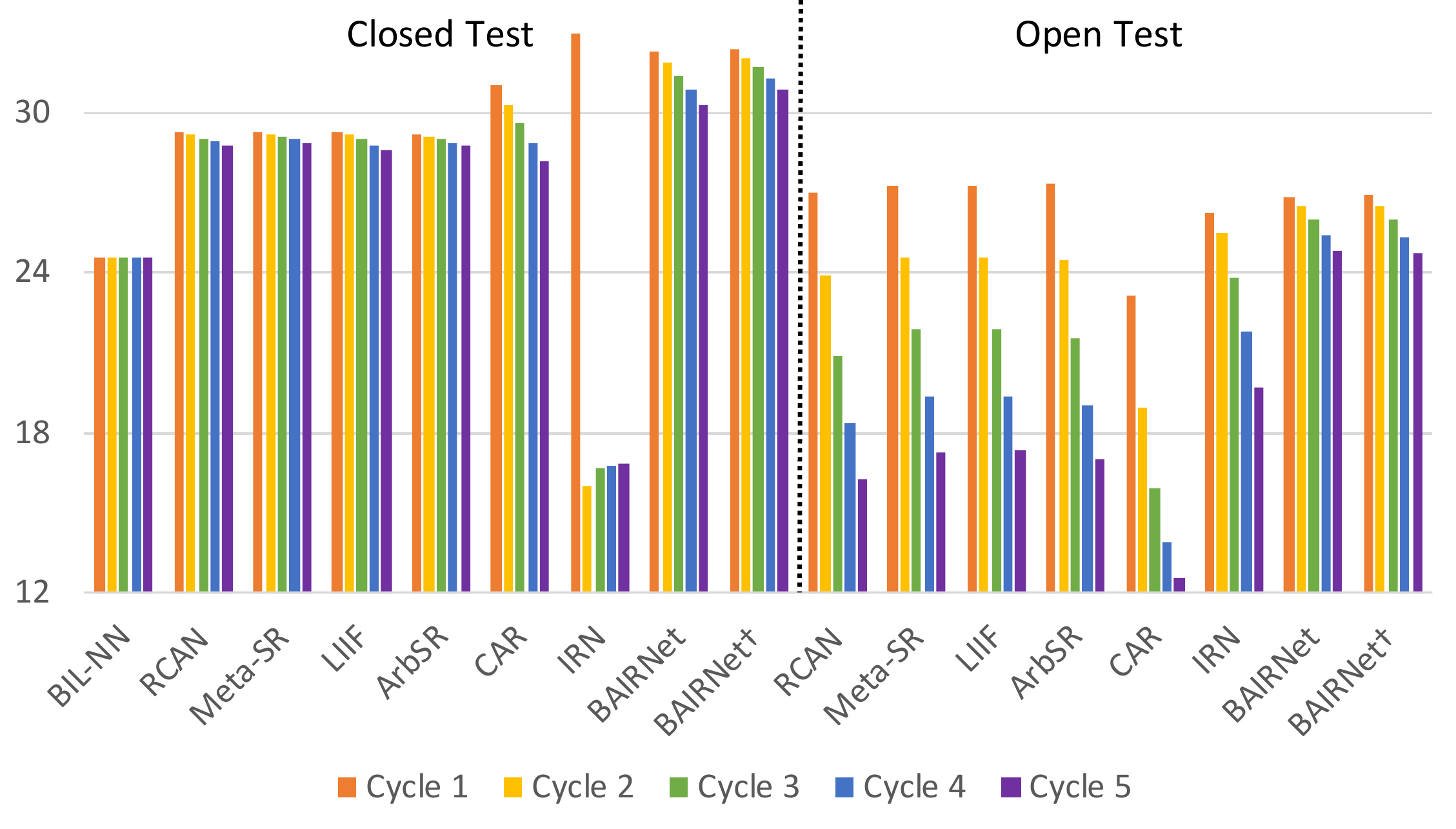}
 \end{center}
 \vspace{-5pt}
 \caption{PSNR results from closed and open idempotence tests for 1-5 cycles on DIV2K validation set ($\times 4$).}
 \vspace{-5pt}
 \label{fig:idem}
 \end{figure}

To avoid extra interpolation for RCAN and IRN, a $\times 4$ scale is chosen here
for testing on the DIV2K validation set and the results are compared in Fig.~\ref{fig:idem}.
For the closed test, BIL-NN (bilinear for downsampling and nearest-neighbour for upsampling)
is included as a perfectly idempotent reference.  IRN has the best performance
at cycle 1 with its bidirectional learning and invertible network structure.
However, there is a drastic drop from cycle 2 and hereafter, probably caused by the random
latent variable sampling during the upscaling process.  For our method, both BAIRNet 
and BAIRNet$^\dagger$ are included to show the improvement in idempotence of BAIRNet$^\dagger$,
which is fine-tuned from BAIRNet using the proxy objective of multiple-cycle losses.
At cycle 1, both are closely behind IRN, and their additional quality losses from multiple cycles are gradual.
After 5 cycles, both are still significantly better than upscaling-only models at cycle 1, 
and BAIRNet$^\dagger$ is clearly better than BAIRNet.
This shows the advantage of our method in robustness to repetitive rescaling cycles in closed settings,
and the effectiveness of multi-cycle losses.  For open tests on the right,
while all models are subject to significant performance losses at cycle 1 comparing to closed tests,
our models have a much slower degradation for multiple cycles.
Due to page limitation, more visual examples of cycle idempotence tests are included as supplementary
materials.

 \begin{table}[h]
 \footnotesize
 \setlength{\tabcolsep}{2pt}
     \vspace{-4pt}
     \caption{PSNR improvements over base BAIRNet after fine-tuning using $N$-cycle losses (Eq.~\ref{eq:optm}), testing 5 cycles each for 3 scales.}
     \vspace{-10pt}
 \begin{center}
 \begin{tabular}{cccccccccc}
 \hline \hline
 \multirow{2}{*}{Cycle} & \multicolumn{3}{c}{$N=1$} & \multicolumn{3}{c}{$N=3$} & \multicolumn{3}{c}{$^{^{\wt{\ddagger}}}$$N=5$} \\
 & $\times 4$ & $\times 3$ & $\times 2$ & $\times 4$ & $\times 3$ & $\times 2$ & $\times 4$ & $\times 3$ & $\times 2$\\
 \midrule
1 & \red{0.09} & \bp{0.09} & 0.09 & \bp{0.08} & \red{0.10} & \bp{0.13} & 0.05 & 0.08 & \red{0.13}\\
2 & 0.14 & 0.16 & 0.21 & \red{0.16} & \bp{0.21} & \bp{0.33} & \bp{0.15} & \red{0.21} & \red{0.34}\\
3 & 0.21 & 0.24 & 0.32 & \bp{0.27} & \bp{0.36} & \bp{0.56} & \red{0.28} & \red{0.39} & \red{0.60}\\
4 & 0.29 & 0.34 & 0.42 & \bp{0.40} & \bp{0.54} & \bp{0.78} & \red{0.44} & \red{0.60} & \red{0.86}\\
5 & 0.38 & 0.43 & 0.49 & \bp{0.54} & \bp{0.73} & \bp{1.00} & \red{0.61} & \red{0.82} & \red{1.11}\\
 \hline \hline
 \end{tabular}
 \end{center}
 \label{tab:N}
 \vspace{-11pt}
 \end{table}

To study the effectiveness of various $N$ in Eq.~\ref{eq:optm},
the base BAIRNet model is trained for another 200 epochs using $N=1, 3, 5$ respectively.
As shown in Table~\ref{tab:N}, the improvements in PSNR for three fine-tuned
models are compared for 1-5 cycles at 3 different scales.  There
are consistent improvements across scales and cycles even when $N=1$, indicating the base
BAIRNet is not fully trained.  For $N=3$, it is shown to improve PSNR more significantly
after multi-cycles, especially for smaller scales, while only trailing by 0.01 at
1-cycle for $\times 4$.  For $N=5$, the corresponding gain at multi-cycles is larger, but there
is trade-off of accuracy at 1-cycle.
Overall, it is demonstrated that the proposed proxy objective is effective at increasing
model robustness in cycle idempotence while maintaining high performance at the primary
goal of 1-cycle reconstruction accuracy.

 \begin{table*}[h]
 \scriptsize
 \setlength{\tabcolsep}{2pt}
     \vspace{0pt}
     \caption{PSNR results achieved for both symmetric and asymmetric scale factors (methods in \textbf{bold} require extra interpolations).}
     \vspace{-3pt}
 \begin{center}
 \begin{tabular}{lcccccccccccccccccccc}
 \hline \hline
 \multirow{2}{*}{} & \multicolumn{4}{c}{Set5} & \multicolumn{4}{c}{Set14} & \multicolumn{4}{c}{BSD100} & \multicolumn{4}{c}{Urban100} & \multicolumn{4}{c}{$^{^{\wt{\ddagger}}}$Manga109}\\
 & $\frac{\times 2}{\times 3}$ & $\frac{\times 1.6}{\times 3.2}$ & $\frac{\times 3.6}{\times 1.2}$ & $\times 2.5$
 & $\frac{\times 2}{\times 3}$ & $\frac{\times 1.6}{\times 3.2}$ & $\frac{\times 3.6}{\times 1.2}$ & $\times 2.5$
 & $\frac{\times 2}{\times 3}$ & $\frac{\times 1.6}{\times 3.2}$ & $\frac{\times 3.6}{\times 1.2}$ & $\times 2.5$
 & $\frac{\times 2}{\times 3}$ & $\frac{\times 1.6}{\times 3.2}$ & $\frac{\times 3.6}{\times 1.2}$ & $\times 2.5$
 & $\frac{\times 2}{\times 3}$ & $\frac{\times 1.6}{\times 3.2}$ & $\frac{\times 3.6}{\times 1.2}$ & $\times 2.5$ \vspace{1pt}\\
 \midrule
  ArbSR   & 35.90 & 35.90 & 35.85 & 36.21 & 31.89 & 31.96 & 31.59 & 31.99 & 30.58 & 30.87 & 30.24 & 30.51 & 30.59 & 30.60 & 29.74 & 30.68 & 36.17 & 35.88 & 35.30 & 36.67 \\
  \textbf{IRN} & 38.66 & 38.55 & \bp{38.80} & 39.78 & 35.49 & 35.31 & \bp{35.02} & 36.39 & 34.87 & 34.69 & \bp{34.19} & 35.56 & 32.75 & 32.44 & \bp{32.09} & 33.99 & 37.42 & 37.08 & \bp{37.12} & 39.33 \\
 \textbf{BAIRNet} & \bp{39.40} & \bp{39.05} & 38.34 & \bp{40.03} & \bp{36.00} & \bp{35.71} & 34.36 & \bp{36.53} & \bp{35.33} & \bp{35.14} & 33.47 & \red{36.24} & \bp{34.14} & \bp{33.34} & 31.52 & \red{36.51} & \bp{39.01} & \bp{38.30} & 36.95 & \bp{40.11} \\
 BAIRNet$^{\ddagger}$ & \red{40.06} & \red{40.42} & \red{40.11} & \red{40.16} & \red{36.66} & \red{37.00} & \red{36.33} & \red{36.68} & \red{36.28} & \red{36.96} & \red{36.64} & \bp{36.17} & \red{36.32} & \red{36.74} & \red{35.82} & \bp{36.43} & \red{40.01} & \red{40.27} & \red{39.42} & \red{40.14} \\
 \hline \hline
 \end{tabular}
 \end{center}
 \label{tab:xy}
 \vspace{-8pt}
 \end{table*}

\begin{figure*}[t]
\captionsetup[subfigure]{font=footnotesize, labelformat=empty}
\begin{center}
  \begin{subfigure}[b]{0.135\textwidth}
    \centering
      \includegraphics[width=\textwidth, interpolate=false]{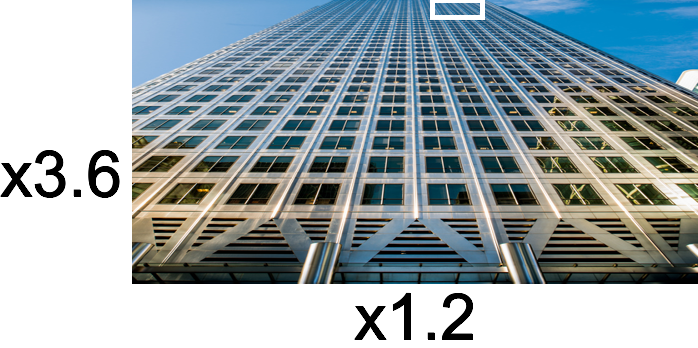}
  \end{subfigure} \hspace*{-0.4em}
  \begin{subfigure}[b]{0.09\textwidth}
    \centering
      \includegraphics[width=\textwidth, interpolate=false]{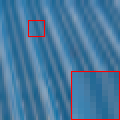}
  \end{subfigure} \hspace*{-0.4em}
  \begin{subfigure}[b]{0.09\textwidth}
    \centering
      \includegraphics[width=\textwidth, interpolate=false]{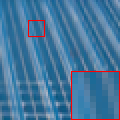}
  \end{subfigure} \hspace*{-0.4em}
  \begin{subfigure}[b]{0.09\textwidth}
    \centering
      \includegraphics[width=\textwidth, interpolate=false]{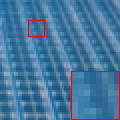}
  \end{subfigure} \hspace*{-0.4em}
  \begin{subfigure}[b]{0.09\textwidth}
    \centering
      \includegraphics[width=\textwidth, interpolate=false]{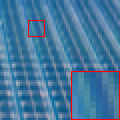}
  \end{subfigure} \hspace*{-0.4em}
  \begin{subfigure}[b]{0.09\textwidth}
    \centering
      \includegraphics[width=\textwidth, interpolate=false]{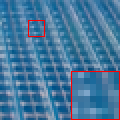}
  \end{subfigure} \hspace*{-0.4em}
  \begin{subfigure}[b]{0.09\textwidth}
    \centering
      \includegraphics[width=\textwidth, interpolate=false]{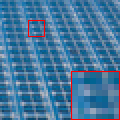}
  \end{subfigure}

  \hspace*{0.81em}
  \begin{subfigure}[b]{0.09\textwidth}
    \centering
      \includegraphics[width=\textwidth, interpolate=false]{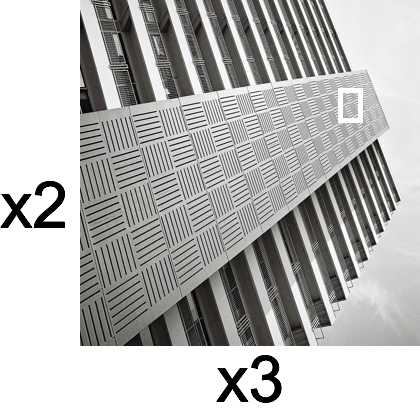}
      \caption{LR\wt{$^{\ddagger}$}}
  \end{subfigure} \hspace*{0.81em}
  \begin{subfigure}[b]{0.09\textwidth}
    \centering
      \includegraphics[width=\textwidth, interpolate=false]{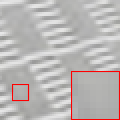}
      \caption{Bicubic\wt{$^{\ddagger}$}}
  \end{subfigure} \hspace*{-0.4em}
  \begin{subfigure}[b]{0.09\textwidth}
    \centering
      \includegraphics[width=\textwidth, interpolate=false]{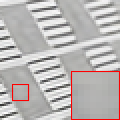}
      \caption{ArbSR\wt{$^{\ddagger}$}}
  \end{subfigure} \hspace*{-0.4em}
  \begin{subfigure}[b]{0.09\textwidth}
    \centering
      \includegraphics[width=\textwidth, interpolate=false]{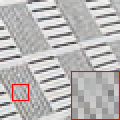}
      \caption{\textbf{IRN}\wt{$^{\ddagger}$}}
  \end{subfigure} \hspace*{-0.4em}
  \begin{subfigure}[b]{0.09\textwidth}
    \centering
      \includegraphics[width=\textwidth, interpolate=false]{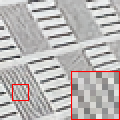}
      \caption{\textbf{BAIRNet}\wt{$^{\ddagger}$}}
  \end{subfigure} \hspace*{-0.4em}
  \begin{subfigure}[b]{0.09\textwidth}
    \centering
      \includegraphics[width=\textwidth, interpolate=false]{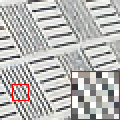}
      \caption{BAIRNet$^{\ddagger}$}
  \end{subfigure} \hspace*{-0.4em}
  \begin{subfigure}[b]{0.09\textwidth}
    \centering
      \includegraphics[width=\textwidth, interpolate=false]{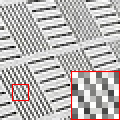}
      \caption{GT\wt{$^{\ddagger}$}}
  \end{subfigure}
 \end{center}
\vspace{-3pt}
    \caption{Visual examples of arbitrary asymmetric rescaling from Urban100 test set (Best viewed for online version).}
\label{fig:xy}
\vspace{-3pt}
\end{figure*}

\begin{table}[h]
	\centering
	\footnotesize
	\setlength{\tabcolsep}{2pt}
	\caption{PSNR results for large out-of-distribution scales.}
	\vspace{3pt}
	\begin{tabular}{lccccc} 
		\hline \hline
		{} & {$\times 6$} & {$\times 12$} & {$\times 18$} & {$\times 24$} & {$\times 30$}\\
		\midrule
		Bicubic & 24.82 & 22.27 & 21.00 & 20.19 & 19.59\\
		LIIF    & \bp{27.02} & \bp{23.95} & \bp{22.40} & \bp{21.40} & \bp{20.71}\\
		ArbSR   & 26.61 & 23.07 & 21.45 & 20.49 & 19.81\\
		BAIRNet & \red{29.29} & \red{25.55} & \red{23.84} & \red{22.75} & \red{21.97}\\
		\hline \hline
	\end{tabular}
\label{tab:big}
\end{table}

\subsection{Out-of-distribution Assessment}
\label{sec:ood}
While our model is trained with symmetric scale factors randomly distributed
between $\times 1- \times4$ and mainly tested using such
in-distribution settings, there is no such limitation in the capacity of the proposed method.
For assessment, as shown in Table~\ref{tab:xy} and \ref{tab:big} respectively,
our model is compared with ArbSR for asymmetric scales and with LIIF for large scales. 
In both cases, BAIRNet is used as is for upscaling
by simply changing the output resolution and using corresponding $\tilde{\phi}$
as in Eq.~\ref{eq:func}.  These out-of-distribution tests further demonstrate robustness of
our proposed method.

For asymmetric scales $\frac{s_v}{s_h}$ where $s_v$ is the vertical scale and $s_h$ is for horizontal,
comparisons of 5 benchmark test sets are shown in Table~\ref{tab:xy}.
Results from ArbSR are included as a SOTA baseline for unidirectional models where only upscaling is le.
For bidirectional IRN, although additional interpolations are needed for both downscaling and upscaling
as only $\times 2$ and $\times 4$ models are available, its performance is far more superior comparing
to the ArbSR baseline.  For our base BAIRNet, pre-interpolation is only needed for the downscaling stage,
where the input GT image is resampled using bicubic interpolation with a $s_m/s_v$ vertical scale
and a $s_m/s_h$ horizontal scale.  Here $s_m = \sqrt{s_h s_v}$, which will convert the asymmetric
scale to symmetric scale of $s_m$ for downscaling, while keeping the number of input pixels
approximately the same as GT for fair comparison.  It is shown in Table~\ref{tab:big}
that BAIRNet is better than IRN with the exception of $\frac{\times 3.6}{\times 1.2}$,
similar to the observation from Fig.~\ref{fig:scale} that IRN is slightly better than
BAIRNet for scales larger than $\times 3.5$.  Unlike IRN that is limited to training data
with symmetric scales, BAIRNet could be further trained using data with asymmetric scales
and the fine-tuned model is denoted as BAIRNet$^{\ddagger}$.
It no longer needs the initial symmetric conversion step and
shows further significant improvements in asymmetric tests, while subjects to slight degradation
in symmetric test for only 2 out 5 test sets.  Visual examples in Fig.~\ref{fig:xy} clearly
show that BAIRNet$^{\ddagger}$ is able to reproduce more fine details at random asymmetric scales.

\begin{figure}[t]
\captionsetup[subfigure]{font=footnotesize, labelformat=empty}
\begin{center}
  \begin{subfigure}[b]{0.15\textwidth}
    \centering
      \includegraphics[width=\textwidth, interpolate=false]{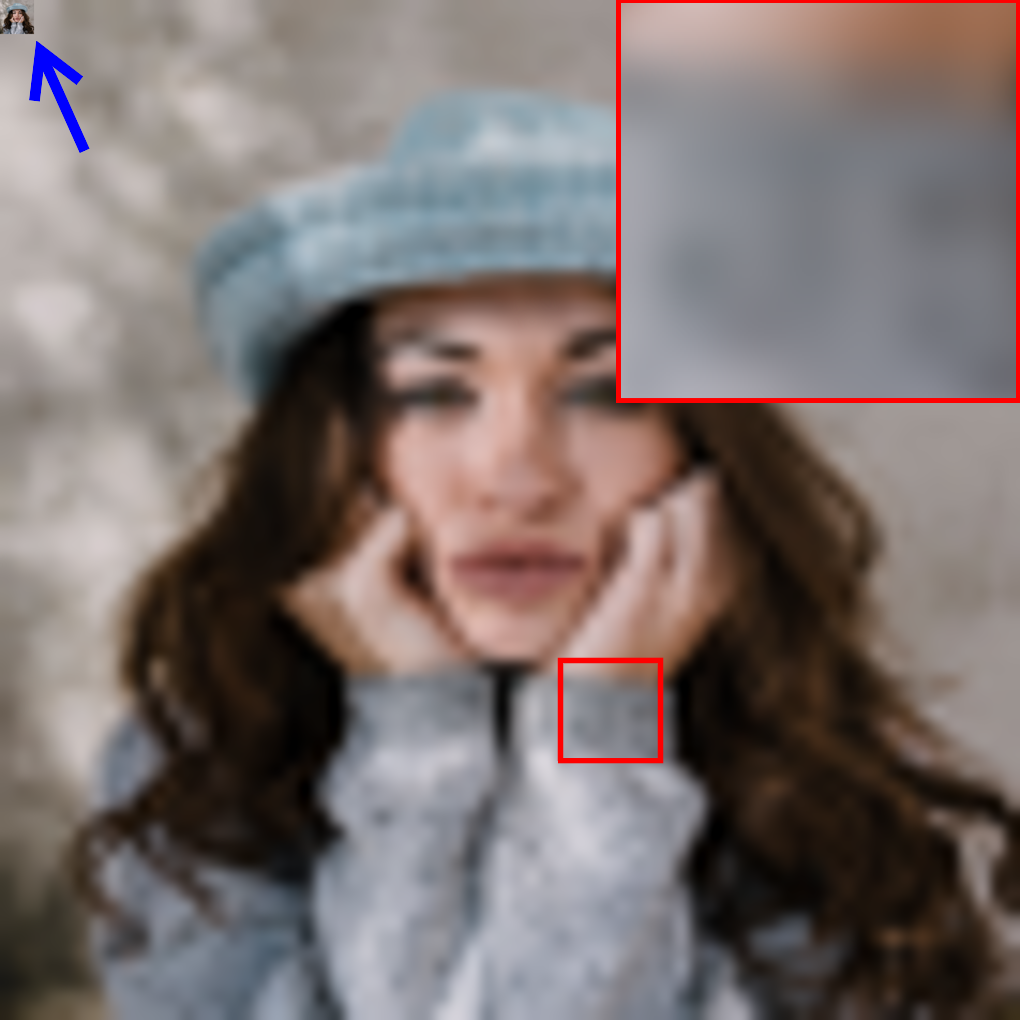}
      \caption{Bicubic}
  \end{subfigure} \hspace*{-0.45em}
  \begin{subfigure}[b]{0.15\textwidth}
    \centering
      \includegraphics[width=\textwidth, interpolate=false]{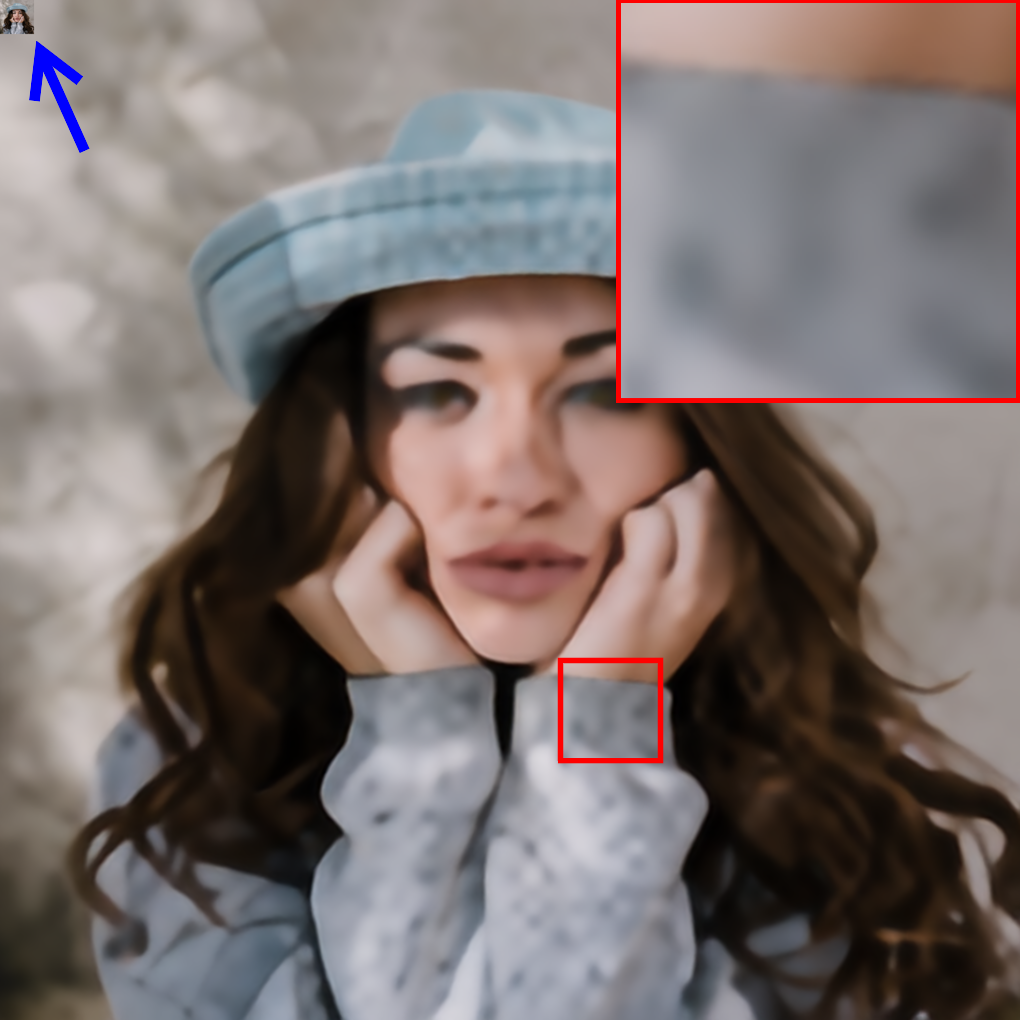}
      \caption{LIIF}
  \end{subfigure} \hspace*{-0.45em}
  \begin{subfigure}[b]{0.15\textwidth}
    \centering
      \includegraphics[width=\textwidth, interpolate=false]{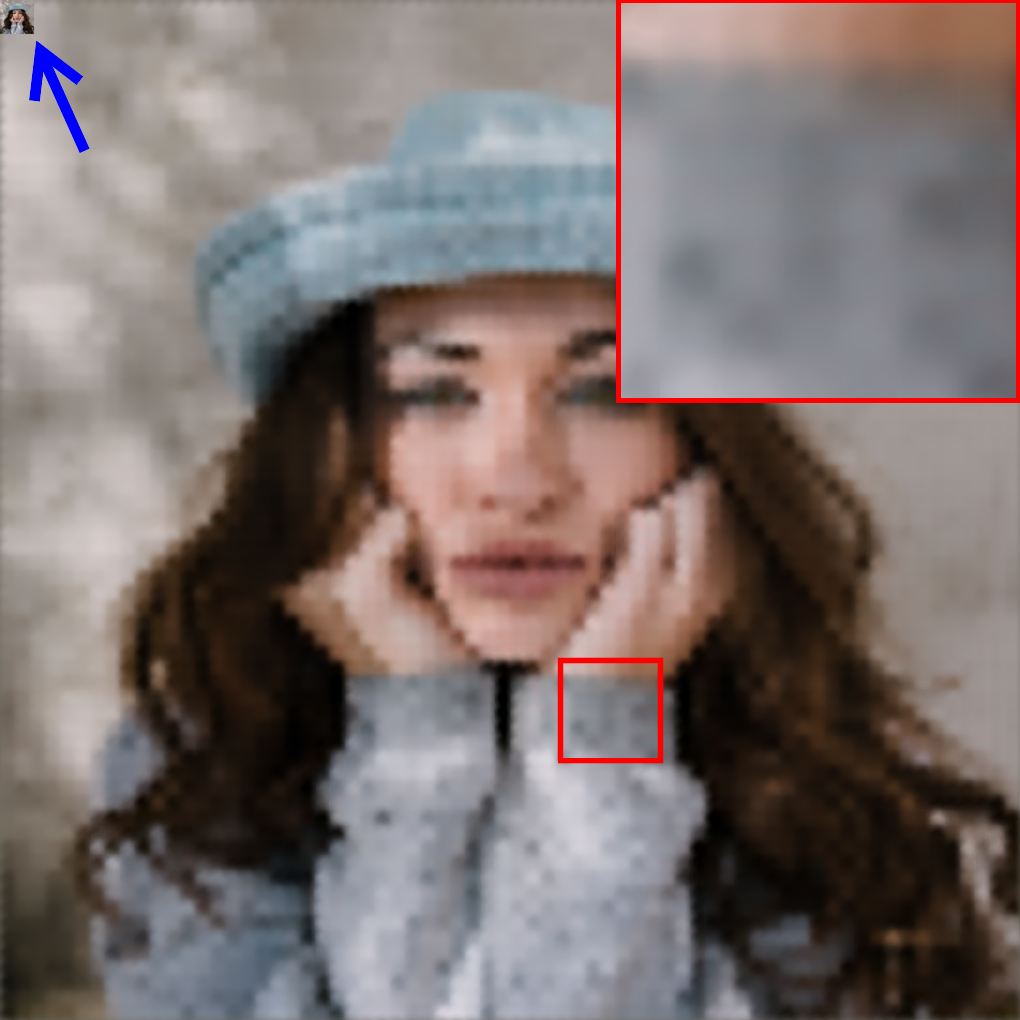}
      \caption{ArbSR}
  \end{subfigure} \hspace*{-0.45em}
  \begin{subfigure}[b]{0.15\textwidth}
    \centering
      \includegraphics[width=\textwidth, interpolate=false]{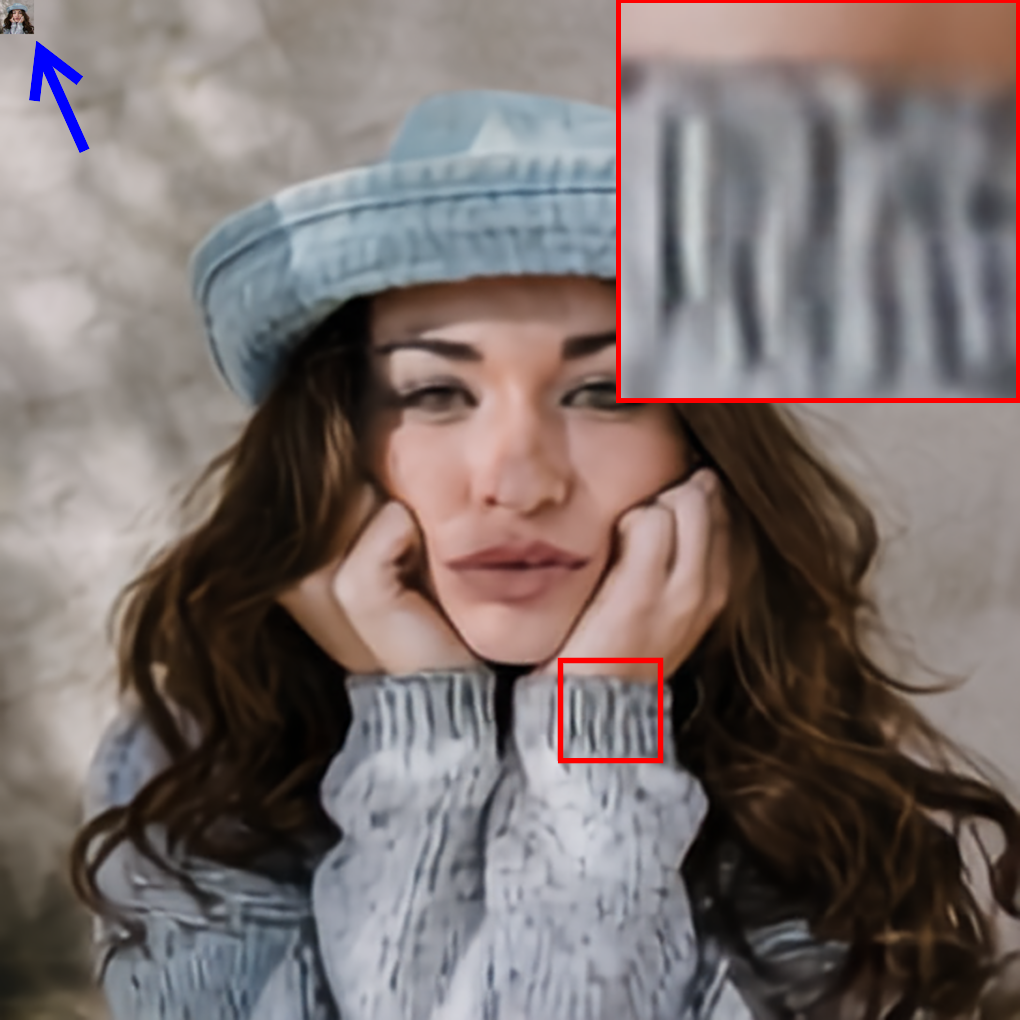}
      \caption{BAIRNet}
  \end{subfigure} \hspace*{-0.45em}
  \begin{subfigure}[b]{0.15\textwidth}
    \centering
      \includegraphics[width=\textwidth, interpolate=false]{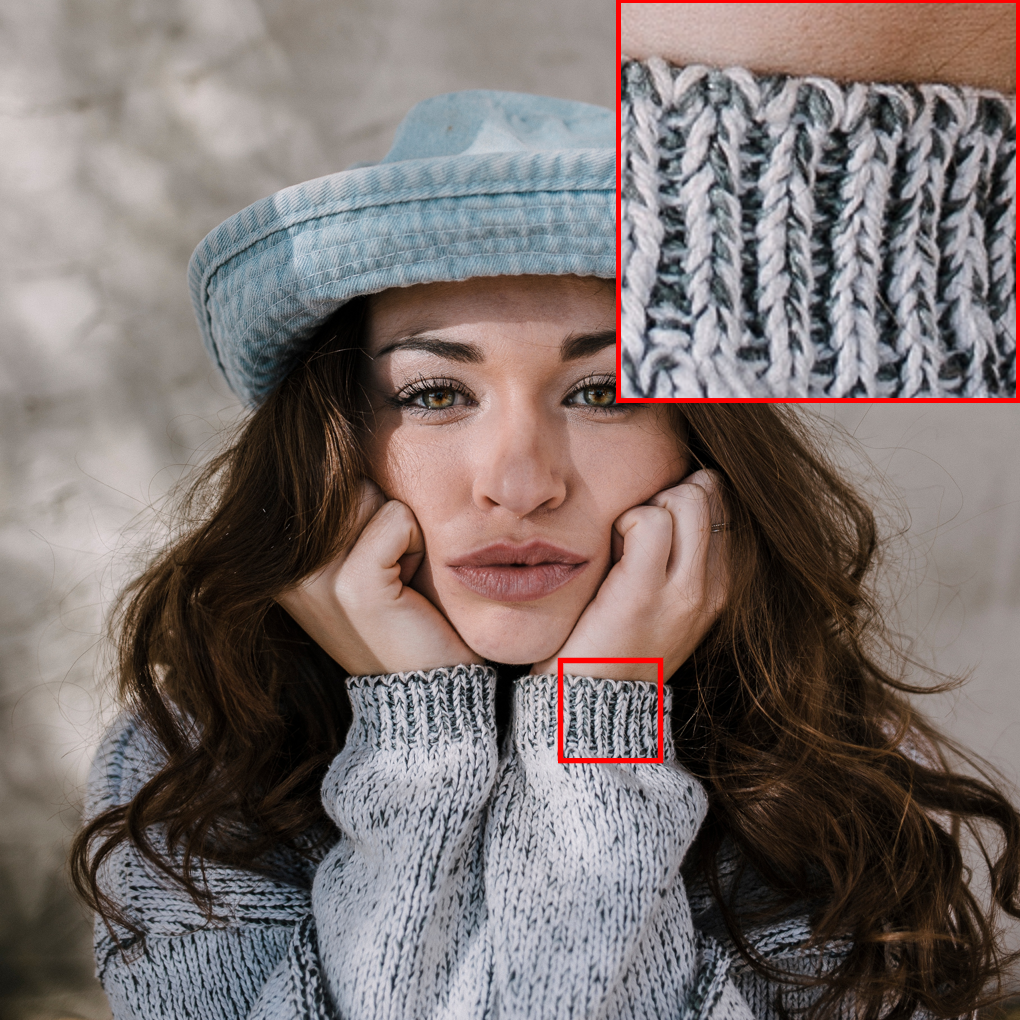}
      \caption{GT}
  \end{subfigure}
 \end{center}
   \vspace{-3pt}
    \caption{Examples of large scale factors ($\times 30$) with \bp{$\uparrow$} pointing to corresponding LR inputs (Best viewed when zooming in).}
   \vspace{-3pt}
\label{fig:big}
\end{figure}

For tests of large scales as shown in Table~\ref{tab:big},
although all models are trained from images with scales up to $\times 4$,
LIIF is much more robust for large scales up to $\times 30$ comparing to ArbSR.
For testing BAIRNet at scale $s$, the GT image is pre-downscaled by a scale of $s/4$
using bicubic so the downscaling step in BAIRNet is capped at $\times 4$.
This is a reasonable choice as this reduces the number of pixels
used for the time-consuming step of downscaling feature encoding.
It is shown that BAIRNet is consistently much better than LIIF quantitatively.
For visual examples in Fig.~\ref{fig:big}, BAIRNet is clearly able to recover
sharper details comparing to the others.
All LR inputs, either from bicubic resizing as in LIIF and ArbSR or
downscaling by BAIRNet, have the same low resolution and are included in Fig.~\ref{fig:big} for reference.

\begin{figure*}[ht]
\captionsetup[subfigure]{font=footnotesize, labelformat=empty}
\begin{center}
  \begin{subfigure}[b]{0.08\textwidth}
    \centering
      \includegraphics[width=\textwidth, interpolate=false]{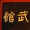}
  \end{subfigure} \hspace*{-0.45em}
  \begin{subfigure}[b]{0.08\textwidth}
    \centering
      \includegraphics[width=\textwidth, interpolate=false]{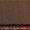}
  \end{subfigure} \hspace*{-0.45em}
  \begin{subfigure}[b]{0.08\textwidth}
    \centering
      \includegraphics[width=\textwidth, interpolate=false]{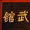}
  \end{subfigure} \hspace*{-0.45em}
  \begin{subfigure}[b]{0.08\textwidth}
    \centering
      \includegraphics[width=\textwidth, interpolate=false]{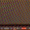}
  \end{subfigure} \hspace*{-0.45em}
  \begin{subfigure}[b]{0.08\textwidth}
    \centering
      \includegraphics[width=\textwidth, interpolate=false]{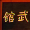}
  \end{subfigure} \hspace*{-0.45em}
  \begin{subfigure}[b]{0.08\textwidth}
    \centering
      \includegraphics[width=\textwidth, interpolate=false]{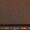}
  \end{subfigure} \hspace*{-0.45em}
  \begin{subfigure}[b]{0.08\textwidth}
    \centering
      \includegraphics[width=\textwidth, interpolate=false]{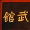}
  \end{subfigure} \hspace*{-0.45em}
  \begin{subfigure}[b]{0.08\textwidth}
    \centering
      \includegraphics[width=\textwidth, interpolate=false]{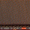}
  \end{subfigure}
  
  \begin{subfigure}[b]{0.161\textwidth}
    \centering
      \includegraphics[width=\textwidth, interpolate=false]{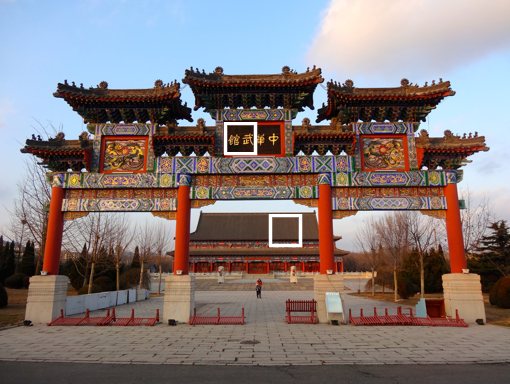}
      \caption{Bicubic\wt{$^{c}$}}
  \end{subfigure} \hspace*{-0.45em}
  \begin{subfigure}[b]{0.161\textwidth}
    \centering
      \includegraphics[width=\textwidth, interpolate=false]{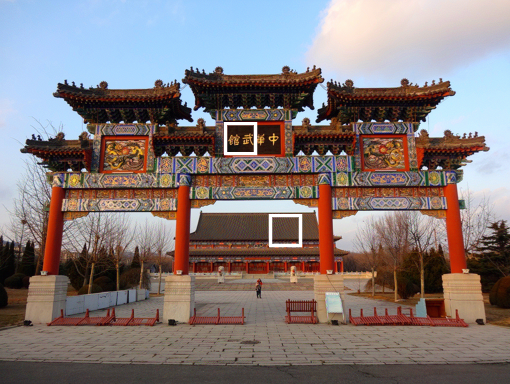}
      \caption{IRN\wt{$^{c}$}}
  \end{subfigure} \hspace*{-0.45em}
  \begin{subfigure}[b]{0.161\textwidth}
    \centering
      \includegraphics[width=\textwidth, interpolate=false]{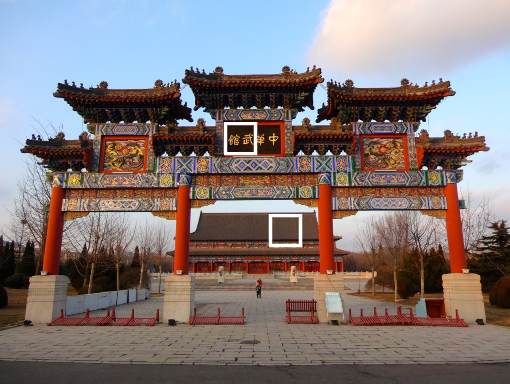}
      \caption{BAIRNet\wt{$^{c}$}}
  \end{subfigure} \hspace*{-0.45em}
  \begin{subfigure}[b]{0.161\textwidth}
    \centering
      \includegraphics[width=\textwidth, interpolate=false]{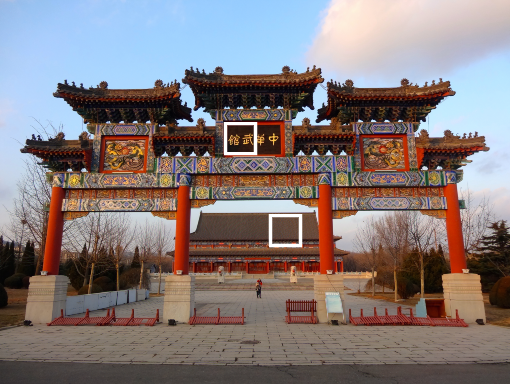}
      \caption{BAIRNet$^{c}$}
  \end{subfigure}
 \end{center}
   \vspace{-12pt}
    \caption{Visual examples of downscaled LR images with false color artifacts in magnified view.}
   \vspace{-12pt}
\label{fig:lr}
\end{figure*}

\subsection{Ablation Study}
To assess the effectiveness of modules like SWF and different LR supervision strategies, the pretrained model
is trained for an additional 200 epoch using different settings as shown in Table~\ref{tab:abl}.
For settings of $\mathcal{L}_{ref}$, $L_2$ means $L_{2}$ loss at pixel level, 
$L^{c}_2$ means $L_2$ for $C_b$ and $C_r$ channels only, and $L^{m}_2$ refers to using mean pixel
value per color channel.  Note that for $L^{c}_2$, $\lambda_2$ is set as 2.  It is shown clearly in
Table~\ref{tab:abl} that the SWF module in downscaling step is beneficial across different scales.
For $\mathcal{L}_{ref}$, it is obvious that weaker supervision leads to improved restoration accuracy overall,
and the weak $L^{m}_2$ is effectively the same as no supervision in LR at all.
For $\mathcal{L}_{rec}$, as images of various scales are included in each
batch of model training, it is expected that those with lower scales have smaller reconstruction
losses naturally.  Intuitively, a simple scale-normalization, $L_s = L_1/s$ where $s$ is the rescaling factor,
is used in the training of BAIRNet.  For comparison, the model is also trained without such
loss normalization, and as shown in the table, it is equivalent with the primary model using $L_s$
for larger scales, but suffers from performance loss at smaller scales.

\begin{table}[h]
	\centering
	\footnotesize
	\setlength{\tabcolsep}{4pt}
	\vspace{0pt}
	\caption{PSNR results for different model and training settings.}
	\vspace{3pt}
	\begin{tabular}{lrcccccc} 
		\hline \hline
		& {$^{^{\wt{\ddagger}}}$D-SWF} & {\xmark} & {\cmark} & {\cmark} & {\cmark} & {\cmark} & {\cmark}\vspace{1pt}\\
		& {$\mathcal{L}_{ref}$} & {$L_2$} & {$L_2$} & {$L^{c}_2$} & {$L^{m}_2$} & {\xmark} & {$L^{m}_2$}\vspace{1.0pt}\\
		& {$\mathcal{L}_{rec}$} & {$L_s$} & {$L_s$} & {$L_s$} & {$L_s$} & {$L_s$} & {$L_1$}\vspace{0.5pt}\\
		\midrule
		\multirow{3}{*}{DIV2K} & $\times 4$ & 31.19 & 31.42 & 31.78 & 32.12 & \bp{32.13} & \red{32.14}\\
    		                   & $\times 3$ & 34.24 & 34.42 & 34.66 & \red{35.13} & \bp{35.12} & 35.10\\
    		                   & $\times 2$ & 38.89 & 39.42 & 39.32 & \bp{40.11} & \red{40.15} & 39.93\\
		\hline \hline
	\end{tabular}
\label{tab:abl}
\vspace{-3pt}
\end{table}

For the specific $L^{c}_2$, it is designed and tested as a measure to suppress false-color artifacts
noticed in downsampled LR from bidirectional models like IRN and ours.
As shown in Fig.~\ref{fig:lr}, the LR images from bicubic downsampling and the
learned bidirectional models like IRN and BAIRNet are hardly differentiable in normal display resolution.
However, when magnified, there are noticeable moir\'{e}-like and false color artifacts
from both models.  Comparing to BAIRNet trained with $L^{m}_2$ in LR,
BAIRNet$^{c}$ trained with the newly designed $L^{c}_2$ loss
is able to successfully suppress false color artifacts as shown
in Fig.~\ref{fig:lr}, while sacrificing the overall restoration quality slightly
as listed in Table~\ref{tab:abl}.

\subsection{Limitations}
The main limitation of our method is its lower speed and large memory consumption for downscaling.
Though it is slightly faster than LIIF in upscaling without using
the feature unfolding option, it slows down both inference and training for downscaling
as the same feature encoder is applied to a larger number of pixels in the HR inputs.
For very large scale factors though, as demonstrated in Section~\ref{sec:ood}, 
it is not necessary to use the full resolution as input and a
pre-downscaling could be applied to reduce number of input pixels and increase efficiency
in downscaling feature encoding.
One potential future improvement is to optimize the feature encoding module in downscaling
for higher efficiency while maintaining accuracy.  Lastly, only the default RDN backbone is used in our model.
A RCAN backbone is expected to further improve performance, as demonstrated in ArbSR, 
but not tested here as it is slower than RDN.

\section{Conclusion}
Current deep learning based image SR and arbitrary SR models are all subject to one
or multiple limiting factors in related to downscaling degradation kernel and
scale factors.  Modeling arbitrary downscaling and upscaling as one
unified subpixel splitting and merging process, a bidirectional arbitrary
image rescaling network (BAIRNet) is shown to improve upscaling accuracy
significantly by jointly optimizing arbitrary upscaling and downscaling.
Cycle idempotence tests are also used to test robustness of
various models when the downscaling-to-upscaling cycle is applied multiple
times, including closed test where the downscaling is limited to model
assumptions or training settings, and open test where the downscaling is
not limited.   For closed and open tests overall,
BAIRNet is the best with great performance at cycle 1
and no sudden drop in accuracy for following cycles.
Additionally, a proxy objective that minimize multi-cycle losses
is demonstrated to further improve model robustness in cycle idempotence.
It is also shown that, even when BAIRNet is only trained for
random symmetric scales between $\times 1\!-\!4$,
it achieves impressive results for rescaling at asymmetric or large
scales, outperforming SOTA methods LIIF and ArbSR
with substantial margins.

\bibliographystyle{unsrt}

\end{document}